\documentclass{iopart}
\usepackage{iopams,url,graphicx,lineno,multirow}  

\begin{document}


\title[Monte Carlo simulation of the SNO proportional counters]{A Monte
  Carlo simulation of the {S}udbury {N}eutrino {O}bservatory proportional
  counters}

\renewcommand{\thefootnote}{\alph{footnote}}

\author{B~Beltran$^1$,
H~Bichsel$^2$,
B~Cai$^{3}$,
G\,A.~Cox$^{2,12}$,
H~Deng$^4$,
J~Detwiler$^5$,
J\,A~Formaggio$^{2,6}$,
S~Habib$^1$,
A\,L.~Hallin$^1$,
A~Hime$^7$,
M~Huang$^{8,9}$,
C~Kraus$^{3,9}$,
H\,R.~Leslie$^3$,
J\,C.~Loach$^{10,5}$,
R~Martin$^{3,5}$,
S~McGee$^2$,
M\,L~Miller$^{6,2}$,
B~Monreal$^{6,13}$,
J~Monroe$^6$,
N\,S~Oblath$^{2,6}$,
S\,J\,M~Peeters$^{10,14}$,
A\,W\,P~Poon$^5$,
G~Prior$^{5,15}$,
K~Rielage$^{2,7}$,
R\,G\,H~Robertson$^2$,
M\,W\,E~Smith$^{2,16}$,
L\,C~Stonehill$^{2,7}$,
N~Tolich$^2$,
T~Van Wechel$^2$,
H~Wan~Chan~Tseung$^{10,2}$,
J~Wendland$^{11}$,
J\,F~Wilkerson$^{2,17}$ and 
A~Wright$^{3,18}$}

\address{$^1$ Department of Physics, University of 
Alberta, Edmonton, Alberta, T6G 2R3, Canada}
\address{$^2$ Center for Experimental Nuclear Physics and Astrophysics, 
and Department of Physics, University of Washington, Seattle, WA 98195, USA}
\address{$^3$ Department of Physics, Queen's University, 
Kingston, Ontario K7L 3N6, Canada}
\address{$^4$ Department of Physics and Astronomy, University of 
Pennsylvania, Philadelphia, PA 19104-6396, USA}
\address{$^5$ Institute for Nuclear and Particle Astrophysics and 
Nuclear Science Division, Lawrence Berkeley National Laboratory, Berkeley, CA 94720, USA}
\address{$^6$ Laboratory for Nuclear Science, Massachusetts Institute of Technology,
 Cambridge, MA 02139, USA}
\address{$^7$ Los Alamos National Laboratory, Los Alamos, NM 87545, USA}
\address{$^8$ Department of Physics, University of Texas at Austin, Austin, TX 78712-0264, USA}
\address{$^9$ Department of Physics and Astronomy, Laurentian 
University, Sudbury, Ontario P3E 2C6, Canada}
\address{$^{10}$ Department of Physics, University of Oxford, 
Denys Wilkinson Building, Keble Road, Oxford OX1 3RH, UK}
\address{$^{11}$ Department of Physics and Astronomy, University of 
British Columbia, Vancouver, BC V6T 1Z1, Canada}
\address{$^{12}$ Current address: Karlsruhe Institute of Technology, D-76021 Karlsruhe, Germany}
\address{$^{13}$ Current address: Department of Physics, University of California at Santa Barbara, La Jolla, CA 93106, USA}
\address{$^{14}$ Current address: Department of Physics and Astronomy, University of Sussex, Brighton BN1 9QH, UK}
\address{$^{15}$ Current address: CERN, Geneva, Switzerland}
\address{$^{16}$ Current address: Department of Astronomy and Astrophysics, The Pennsylvania State University, University Park, PA 16802, USA}
\address{$^{17}$ Current address: Department of Physics, University of North Carolina, Chapel Hill, NC 27599, USA}
\address{$^{18}$ Current address: Physics Department, Princeton University, Princeton, NJ 08544, USA}

\ead{nsoblath@mit.edu}

\begin{abstract}

The third phase of the Sudbury Neutrino Observatory (SNO) experiment added an array of $^3{\rm He}$ proportional counters to the detector. The purpose of this Neutral Current Detection (NCD) array was to observe neutrons resulting from neutral-current solar neutrino-deuteron interactions. We have developed a detailed simulation of the current pulses from the NCD array proportional counters, from the primary neutron capture on $^3{\rm He}$ through the NCD array signal-processing electronics. This NCD array Monte Carlo simulation was used to model the alpha-decay background in SNO's third-phase  $^8{\rm B}$ solar-neutrino measurement.

\end{abstract}


\submitto{\NJP}



\section{The Sudbury Neutrino Observatory}

The Sudbury Neutrino Observatory (SNO) was a heavy-water Cherenkov neutrino detector~\cite{ref:sno_nim} located in the INCO (now Vale) Creighton Mine near Sudbury, Ontario, Canada. The overburden of the experiment is 5890~meters-water-equivalent~\cite{ref:sno_muons}. The solar neutrino target consisted of 1000 tonnes of heavy water (D$_{2}$O) contained in a 12-m-diameter spherical acrylic vessel. An array of 9456 inward-looking photomultiplier tubes (PMTs), supported by a stainless steel geodesic sphere, was used to detect the Cherenkov light produced by the recoil electrons coming from neutrino interactions.  The SNO experiment was operated from 1999 to 2006; the third of three phases of the experiment took place between 2004 and 2006, accumulating 385.17 live days of data.

The SNO experiment was the first neutrino detector capable of detecting the total $^8$B solar-neutrino flux above 2.22~MeV. SNO was designed to provide direct evidence of solar neutrino flavor change by comparing the observed rates of three different reactions~\cite{ref:herbchen}: 
\begin{equation}\label{eq:n_flux}
\begin{array}{lclc}
\nu_x + e^- & \rightarrow & \nu_x + e^- & \ \rm{(ES)} \\
\nu_e + d & \rightarrow & p + p + e^- - 1.44 \;\rm{MeV} & \ \rm{(CC)} \\
\nu_x + d & \rightarrow & p + n + \nu_x - 2.22 \;\rm{MeV} & \ \rm{(NC)}
\end{array}
\end{equation}

The elastic scattering (ES) of neutrinos and electrons is common to all water Cherenkov detectors. It is primarily sensitive to electron neutrinos and the direction of the scattered electron is strongly correlated with the direction of the incoming neutrino. 

Electron neutrinos can interact with deuterons via a charged-current (CC) interaction. It takes place exclusively for electron neutrinos, and the energy of the outgoing electron is understood as a function of the energy of the incoming neutrino.

The third reaction is the neutral-current (NC) breakup of a deuteron. It is equally sensitive to all neutrino flavors, and therefore provides a direct measurement of the total active flux of $^8{\rm B}$ solar neutrinos above an energy threshold of 2.22~MeV. 

The SNO experiment consisted of three phases. During each phase a different method was used to detect the neutron liberated in the NC reaction. In the first phase, neutron capture by deuterons was detected by observing the single 6.13-MeV $\gamma$ emitted. The results were reported in~\cite{ref:sno_d2o,ref:Ahmad:2002jz,ref:Ahmad:2002ka,ref:sno_longd2o}.

For the second phase, 2~tonnes of salt (NaCl) were added to the heavy water.  Neutron capture on $^{35}$Cl significantly increased the neutron detection efficiency because of the larger neutron-capture cross-section and Q value (8.6~MeV).  The emitted $\gamma$ cascade improved the statistical separation between the CC and NC events. Results from phase two were reported in~\cite{ref:sno_salt,ref:sno_nsp}, and a combined analysis of the data from phases one and two was reported in~\cite{ref:sno_leta}. 

In the third and final phase of SNO, an array of $^3$He proportional counters, known as the Neutral Current Detection (NCD) array, was installed in the D$_{2}$O~\cite{ref:ncd_nim}. A summary of the proportional-counter system will be given in the following section. The NCD array allowed SNO to measure the NC flux independent of the ES and CC fluxes. The results for the third phase are reported in~\cite{ref:ncd_prl}.

\section{The Neutral Current Detection Array}\label{sec:ncds}

Based on the pp-chain neutrino fluxes predicted by the Standard Solar Model (SSM), neutrino cross sections in D$_2$O and target properties, approximately 10 neutrons per day would be produced by NC interactions inside the SNO detector.  To successfully measure this small signal, it was essential that the background rates were extremely small and well determined.  Proportional counters are well-suited for low-background measurements: the active detection medium is a gas, which can be highly purified.  Therefore the primary sources of backgrounds are the materials that make up the body of the counters, and contaminants produced cosmogenically or introduced from external sources.  The NCD array was designed to be an efficient, low-background method of detecting free neutrons in the D$_{2}$O~\cite{ref:ncd_nim}.

Neutrons were captured on $^3$He via the following reaction: 
\begin{equation}\label{eq:ncaptonhe}
{}^3\rm{He} + n \rightarrow p + {}^3\rm{H} + 764 \;\rm{keV}.
\end{equation}
The proton ($p$) carried 573~keV of kinetic energy and the triton ($^3$H, or $t$ elsewhere) carried 191~keV. The energetic ions created electron-ion pairs as they deposited energy in the gas. The electrons drifted towards the anode. In the high-electric-field region near the anode they initiated an avalanche of secondary ionization, multiplying the amount of charge collected on the anode by a factor of approximately 220. 

\begin{figure}[htbp]
\centering
\includegraphics[width=0.8\textwidth]{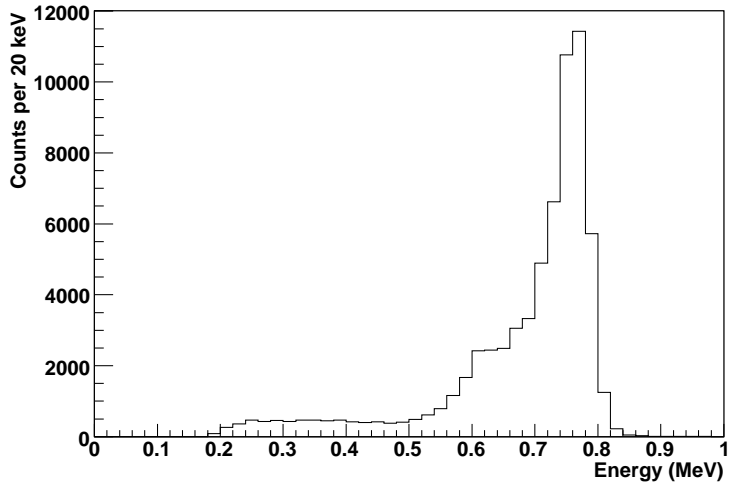}
\caption{NCD array neutron-capture spectrum from a uniformly-distributed $^{24}$Na calibration. The neutron peak is clearly visible at 764~keV and corresponds to the deposition of the full kinetic energy of the proton and triton in the active volume of the NCD counter. The 573-keV shoulder is due to events where the triton energy is fully absorbed by the wall, though it is not visible due to the much larger shoulder that is a result of the space-charge effect discussed in \sref{section:gain}. The 191-keV shoulder is caused by total absorption of the proton energy in the wall.}
\label{fig:ncd_spec}
\end{figure}

The NCD array consisted of 40 strings of 5-cm-diameter cylindrical proportional counters.  Each string was made of three or four electrically-continuous individual counters placed end-to-end.  The strings were between 9 and 11~m in length and were distributed on a 1-m square grid in the D$_2$O volume.  Thirty-six strings were filled with $^3$He and were sensitive to neutrons, while four strings were filled with $^4$He and acted as background control strings with no sensitivity to neutrons.  The shape of the neutron-capture energy spectrum was primarily a result of the geometry of the counters: the proton or triton could hit the wall before depositing all of its energy in the gas. \Fref{fig:ncd_spec} shows the energy spectrum obtained during calibration of the NCD array with a neutron source. A more detailed description of the NCD array is given in~\cite{ref:ncd_nim}.

One of the primary methods for performing neutron calibrations of the NCD array was a $^{24}$Na source uniformly distributed in the D$_2$O volume~\cite{ref:24na}.  The 2.75-MeV $\gamma$ emitted from each $^{24}$Na decay could disintegrate a deuteron, releasing a free neutron into the D$_2$O.  Two such calibrations were performed during SNO's third phase.  Another method for performing neutron calibrations was the use of an encapsulated $^{241}$Am$^9$Be source that could be moved to different positions within the detector.

During operation the copper anode wires were held at 1950~V relative to the counter bodies. The gas consisted of an 85:15 mixture (by pressure) of $^3$He and CF$_4$, with a total gas pressure for each NCD counter of $2.50 \pm 0.01$~atm.  $^3$He is a well-known target for detecting thermalized neutrons~\cite{ref:he3}.  The thermal-neutron-capture cross-section (5333 b~\cite{ref:n_cross}) is seven orders of magnitude larger than that of deuterium.  The CF$_4$ component of the gas mixture acted as quenching gas and reduced the effects of the proton or triton hitting the counter wall.

The NCD array signals were read out only from the tops of the strings since they were submerged vertically in the D$_2$O.  The bottoms of the strings were open circuits to reflect the downward-going portion of current pulses.  A 90-ns delay line at the bottom of the strings further separated the direct and reflected pulses to help in determining the location of the neutron capture along the axis of the NCD string.

The primary backgrounds to the neutron-capture signal were alpha particles emitted by radioactive contaminants within and on the surfaces of the counter bodies and internal parts. Therefore the necessity of maintaining low backgrounds placed certain restrictions on the design and construction of the proportional counters~\cite{ref:ncd_nim}.  Low-background materials were used in all stages of construction, and they were manufactured, treated, and stored carefully to avoid unwanted radon contamination and cosmogenically-created backgrounds.  The $^{232}$Th and $^{238}$U contaminations in the entire NCD array were measured to be less than the goals of 0.5 $\mu$g and 3.8 $\mu$g, respectively.  Another major background component was Rn daughters deposited on surfaces, in particular $^{210}$Po on the counter body walls and other surfaces.  Despite efforts to remove the contamination during NCD array construction, some surface contamination remained: the average $^{210}$Po alpha rate was about 2 alphas/m$^2$-day over the entire 63.52~m$^2$ of the NCD array.

\begin{figure}[htbp]
\centering
\includegraphics[angle=0,width=0.8\textwidth]{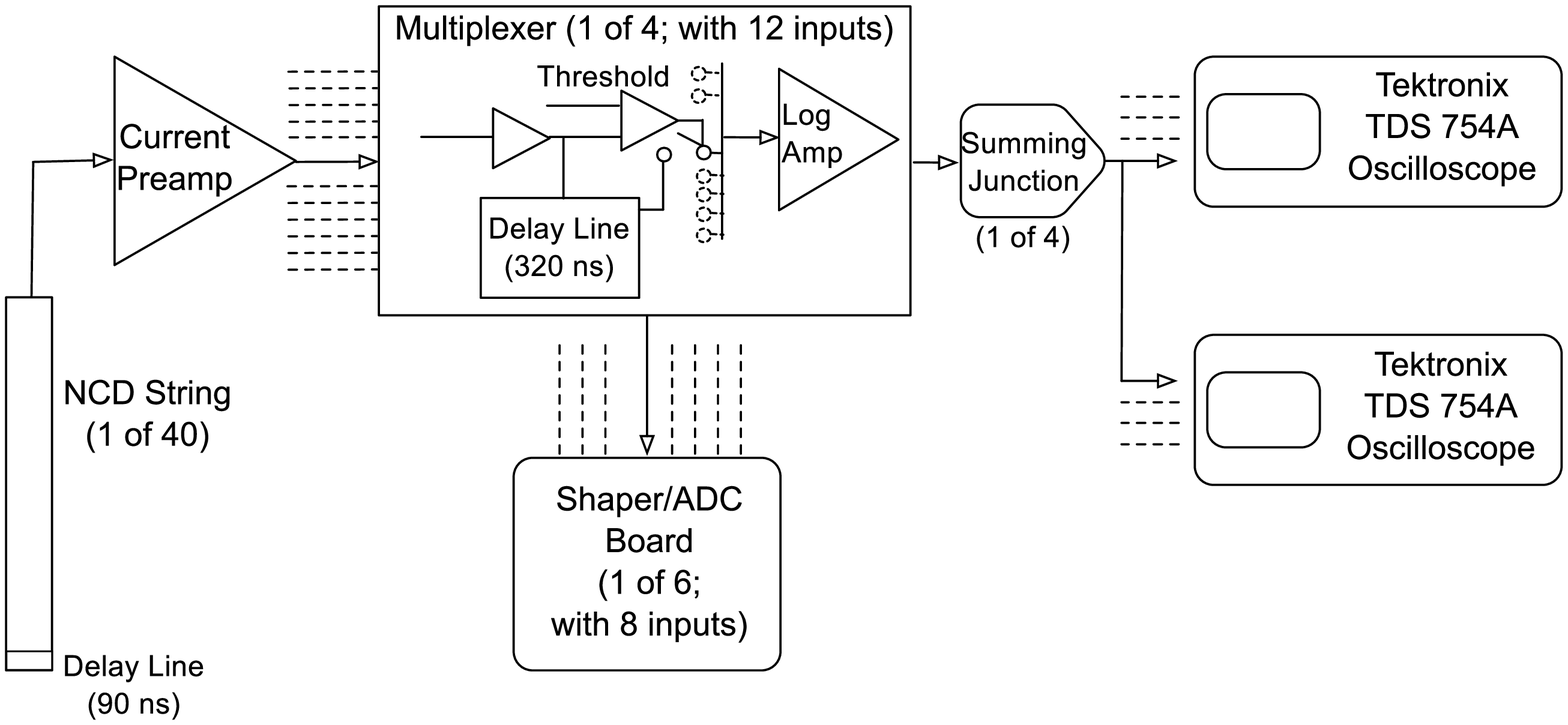}
\caption{Schematic of the NCD array electronics and readout system.}
\label{fig:ncd_elec}
\end{figure}

The NCD array electronics consisted of two independently-triggered readout systems: the ``MUX'' path that digitized the current pulses, and the ``Shaper'' path that integrated the pulses to measure the total energy deposited~\cite{ref:elec}. The basic schematic for the electronics and data acquisition system is shown in \fref{fig:ncd_elec}.

The MUX system recorded the full current pulses from ionization events in the counters with either of two digitizing oscilloscopes. It allowed the use of offline pulse-shape discrimination between neutron-capture signal pulses, and alpha and instrumental background pulses. The MUX path was limited to event rates of several Hz, which was sufficient for solar-neutrino data taking. In contrast, the Shaper path measured only the total charge of the detected events using a shaping/peak-detection network. This system could acquire data at an event rate of several kHz, allowing the measurement of high-rate calibrations, or a galactic supernova, should one have occurred, without being hampered by the system dead-time.

An accurate simulation of the NCD array current pulses is essential for the understanding and elimination of backgrounds in the $^8$B solar-neutrino analysis; the available calibrations of the backgrounds were not sufficient to fully characterize the backgrounds in the data.  The following sections describe our simulation of the physics processes and electronics response that contributed to the characteristics of the observed ionization pulses.

\section{Simulation of Pulses in the NCD Array}
The calculation of current waveforms from events in the NCD system consists of three parts: (1) propagation of the ionizing particle and resulting electrons in the NCD counter gas, (2) calculation of its induced current on the anode, and (3) evaluation of the effects of NCD array hardware on the pulse. In the next section we describe a general method for generating pulse shapes, and sections~\ref{section:ioniz}-\ref{section:imob} describe the major issues pertinent to steps (1) and (2).  \Sref{section:electronics} describes the details of step (3).

\subsection{Method}\label{section:method}
All ionization pulses, regardless of the ionizing particle, are calculated using the same procedure. First, the charged particle trajectory is determined, and divided into $N$ small segments of length $l$. Next, the energy deposited in the gas within each segment is computed. The total current resulting from the whole track at time $t$ is a sum of individual currents from all segments $i$, evaluated at time $t$: 
\begin{equation} \label{equation:summedcurrent}
I_{\mathrm{track}}(t)=\sum_{i=1}^{N} G_i n_{\mathrm{pair},i} I_{i}(t-t_0)
\end{equation}
where 
\begin{equation}
  \label{eq:1}
n_{\mathrm{pair},i}=\frac{\mathrm{d}E}{\mathrm{d}x}\frac{l_i}{W}
\end{equation}
is the number of ion pairs in segment $i$, and $W$ is the mean energy required to produce an electron-ion pair (see \sref{section:W}). The energy loss, $dE$, for a given distance traveled, $dx$, is $dE/dx$.  Stopping powers and the generation of trajectories for the different ionizing particles of interest are described in \sref{section:ioniz}.  The gas gain applied to the electrons from segment $i$ is $G_i$; the value of $G_i$ differs from one segment to another because of statistical and space charge effects (see \sref{section:gain}).  The ``start time'' for the current from the $i^{th}$ segment is $t_0$, which is the difference between the drift times (described in \sref{section:drift}) of ionization electrons from the $i^{th}$ segment, and that of the segment closest to the wire.  $I_{i}(t-t_0)$ is the induced current of a positive charge $q_i=\mathrm{e}\cdot n_{\mathrm{pair},i}$ drifting towards the cathode. It is given by \cite{ref:Wilkinson}:
\begin{equation}\label{equation:wilkinson}
I_i(t-t_0)=-\frac{q_i}{2\mathrm{ln}(b/a)}\frac{1}{t-t_0+\tau}
\end{equation} 
where $\tau$ is the ion-drift time constant, which is related to the ion mobility.  It is characteristic of the NCD counter gas, and was determined to be $5.50 \pm 0.14$~ns (\Sref{section:imob}). The NCD counter anode and cathode radii are $a = 25$~$\mu$m and $b = 2.54$~cm, respectively.  $I_{\mathrm{track}}$ is subsequently convolved with the hardware response as described in \sref{section:electronics}.

The contribution of avalanche electrons to $I_{\mathrm{track}}$ is negligible, because the mean radius at which cascade begins is small ($\sim{}2a$). All electrons in a typical avalanche shower are collected within $\sim{}0.6$~ns, during which time the ions are nearly motionless. The electrons and ions are in close proximity during the shower, so the induced charges at the wire are approximately equal and opposite.  Therefore the net induced current during the electron collection time is small and can be neglected; the current is essentially due to the positive ion drift.

\begin{figure}[ht]
  \centering
     \includegraphics[width=0.8\textwidth]{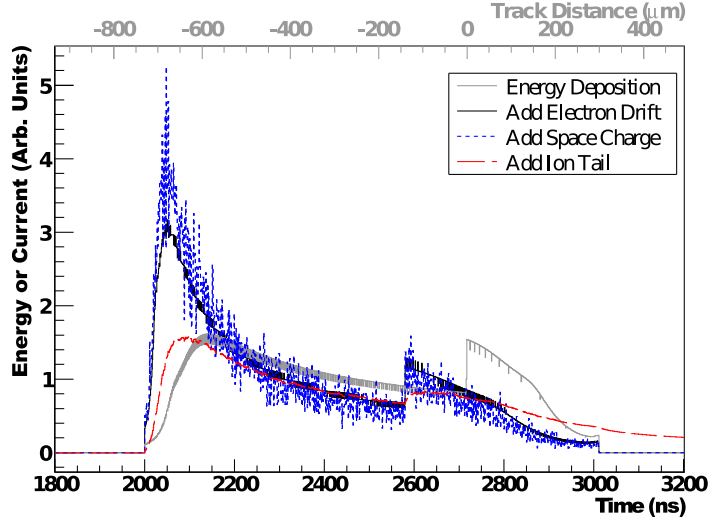}
  \caption{\label{fig:pulsetransform}
Calculation of $I_{\mathrm{track}}$ \eref{equation:summedcurrent} for a $p$-$t$ track perpendicular to the anode, with the proton moving radially inwards. The grey distribution shows the energy deposition per unit of length, ${\rm d}E/{\rm d}x$ for the proton and triton tracks; the track-distance axis is on the top of the plot, with the origin being the start of proton and triton tracks.  The proton travels to the left and the triton to the right.  The peak in the proton energy deposition is the Bragg peak.  The initial triton energy is already below its Bragg peak, so it has a continuously falling distribution.  The black curve is the distribution of electron arrival times at the wire.  This curve appears stretched relative to the energy deposition because of the shape of the electron drift-time distribution and the difference in drift times for each end of the track.  The noise on the first two curves is the result of rebinning the track segments.  The blue short-dashed curve results from applying the gas gains $G_i$, including space-charge effects; the large noise fluctuations are due to avalanche statistics.  The red long-dashed pulse is $I_{\mathrm{track}}$.  The effects of the ion drift are added by convolving \eref{equation:wilkinson} with the previous curve; it effectively acts as a low-pass filter, smoothing out the high-frequency fluctuations and adding a long tail to the pulse. All curves are normalized to the same area.}
\end{figure}

An efficient way of evaluating \Eref{equation:summedcurrent} is to first calculate the distribution of arrival times of electrons at the wire, then convolve that distribution with \Eref{equation:wilkinson} using a fast Fourier transform (FFT) algorithm \cite{ref:FFTW}. The speed of calculation depends, to a large extent, on the number of segments into which a trajectory is divided. The optimal segment length is governed by the frequency of the pulse digitization (1~GHz); it is desirable for electrons from two adjacent segments to reach the wire within one bin width (1~ns).  The restriction on the largest permissible segment size comes from tracks that are perpendicular to the wire and point radially inwards, in which case the segment length should not exceed $4.5$~$\mu$m. It should be also noted that the segment length is roughly comparable with the mean free path of the primary charged particles. For computational accuracy, the size of a segment is chosen to be 1~$\mu$m everywhere along the track, which amounts to $\sim{}11,000$ divisions for a fully-contained 1.1-cm-long proton-triton ($p$-$t$) track.  The various steps in the calculation of $I_{\mathrm{track}}$ are illustrated in \fref{fig:pulsetransform} for a $p$-$t$ track perpendicular to the anode, with the proton moving radially inwards.

\begin{figure}[ht]
  \centering
\includegraphics[width=0.4\textwidth]{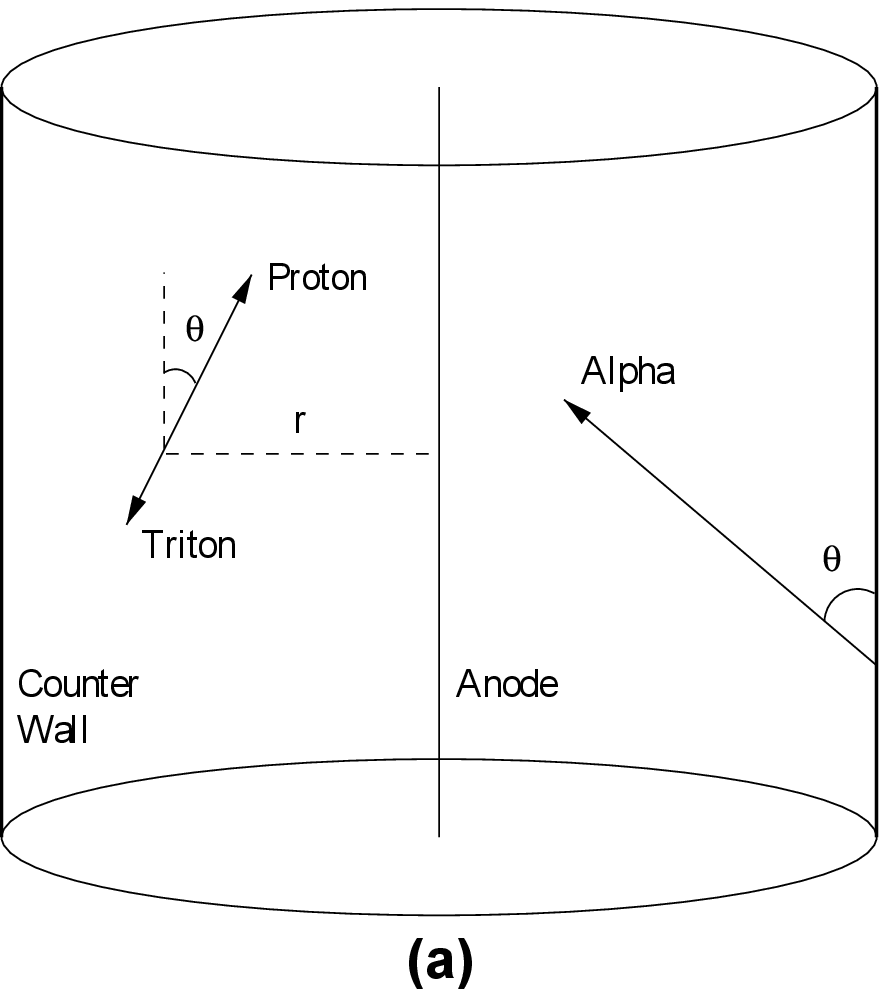}
\includegraphics[width=0.4\textwidth]{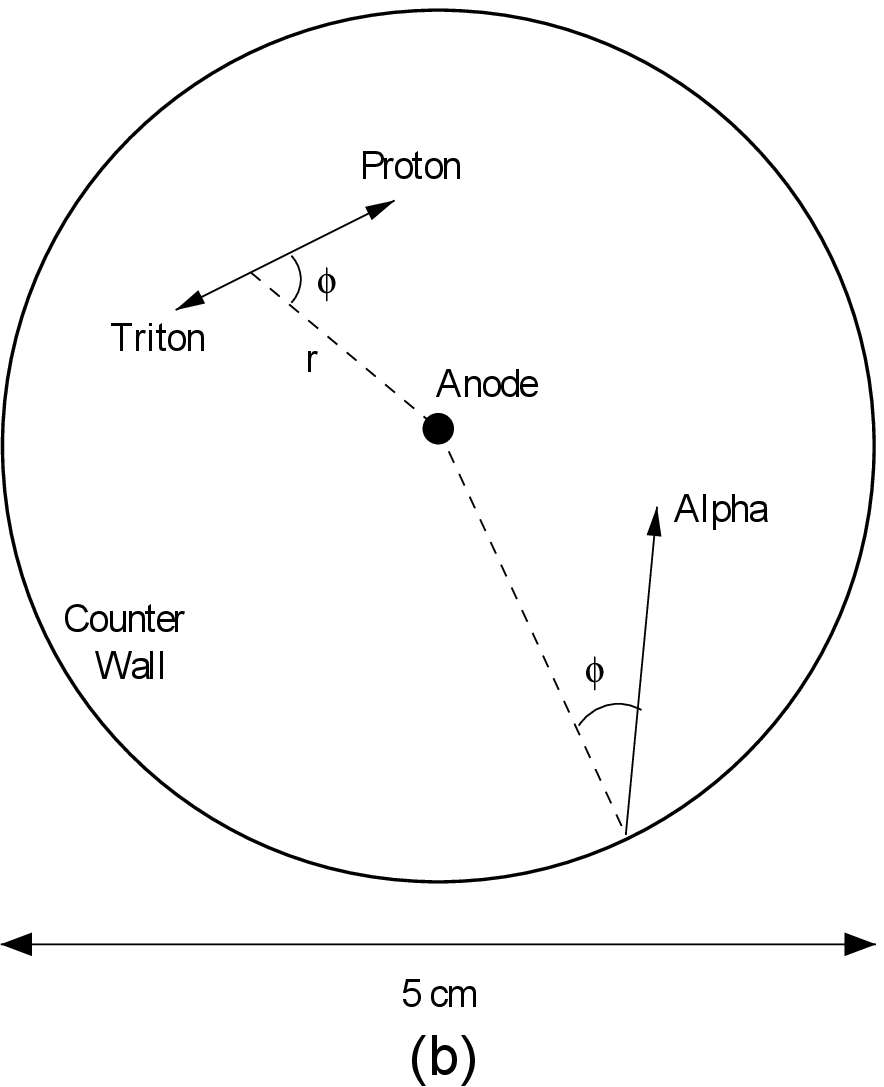}
  \caption{\label{fig:theta_and_phi}
Graphical definition of the angles $\theta$ and $\phi$ used to parameterize the alpha and $p$-$t$ tracks.  $\theta$ is with respect to the $z$ axis, and $r$ always lies in the $x-y$ plane.}
\end{figure}

\begin{figure}[ht]
\includegraphics[width=0.3\textwidth]{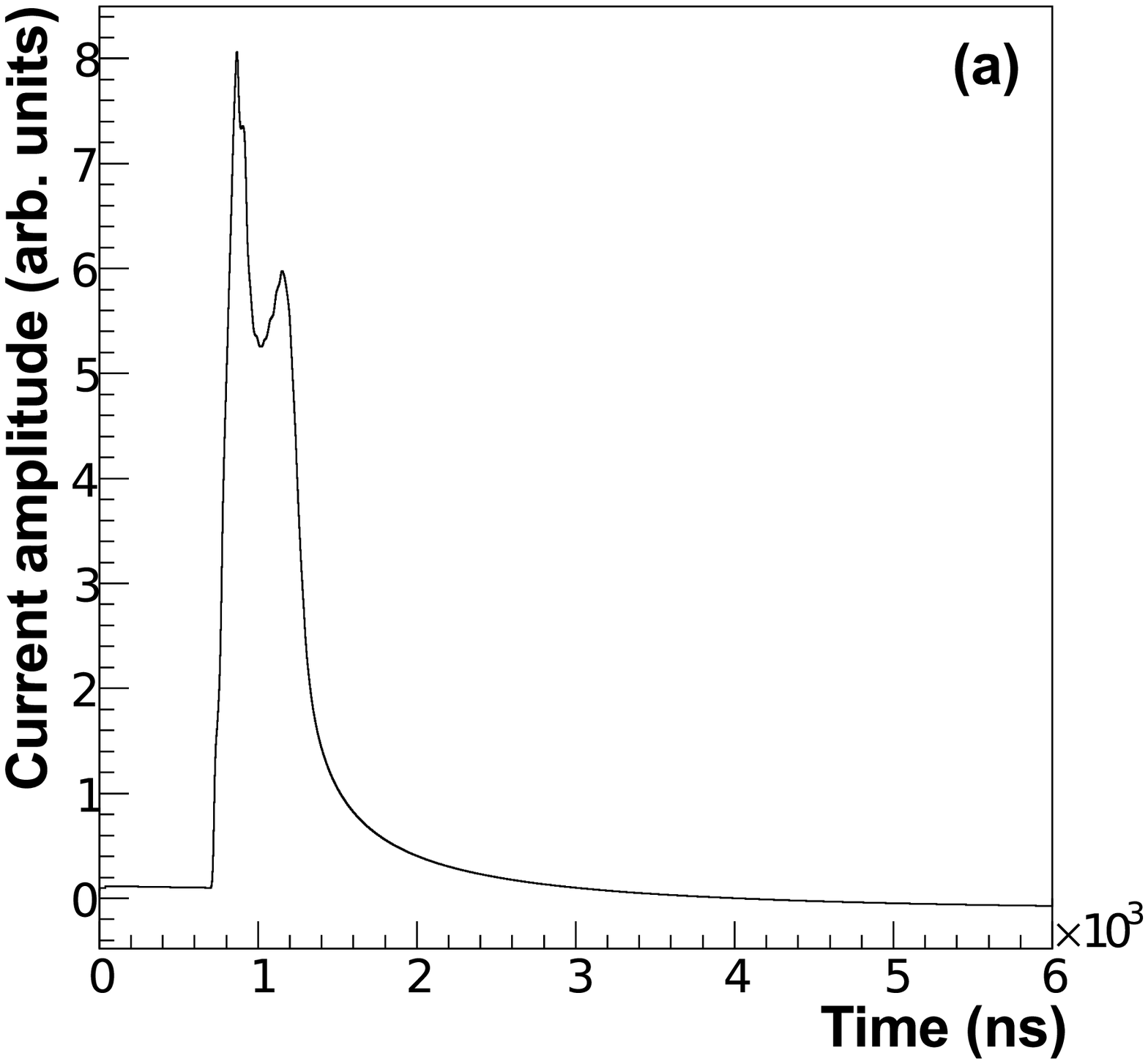}
\includegraphics[width=0.3\textwidth]{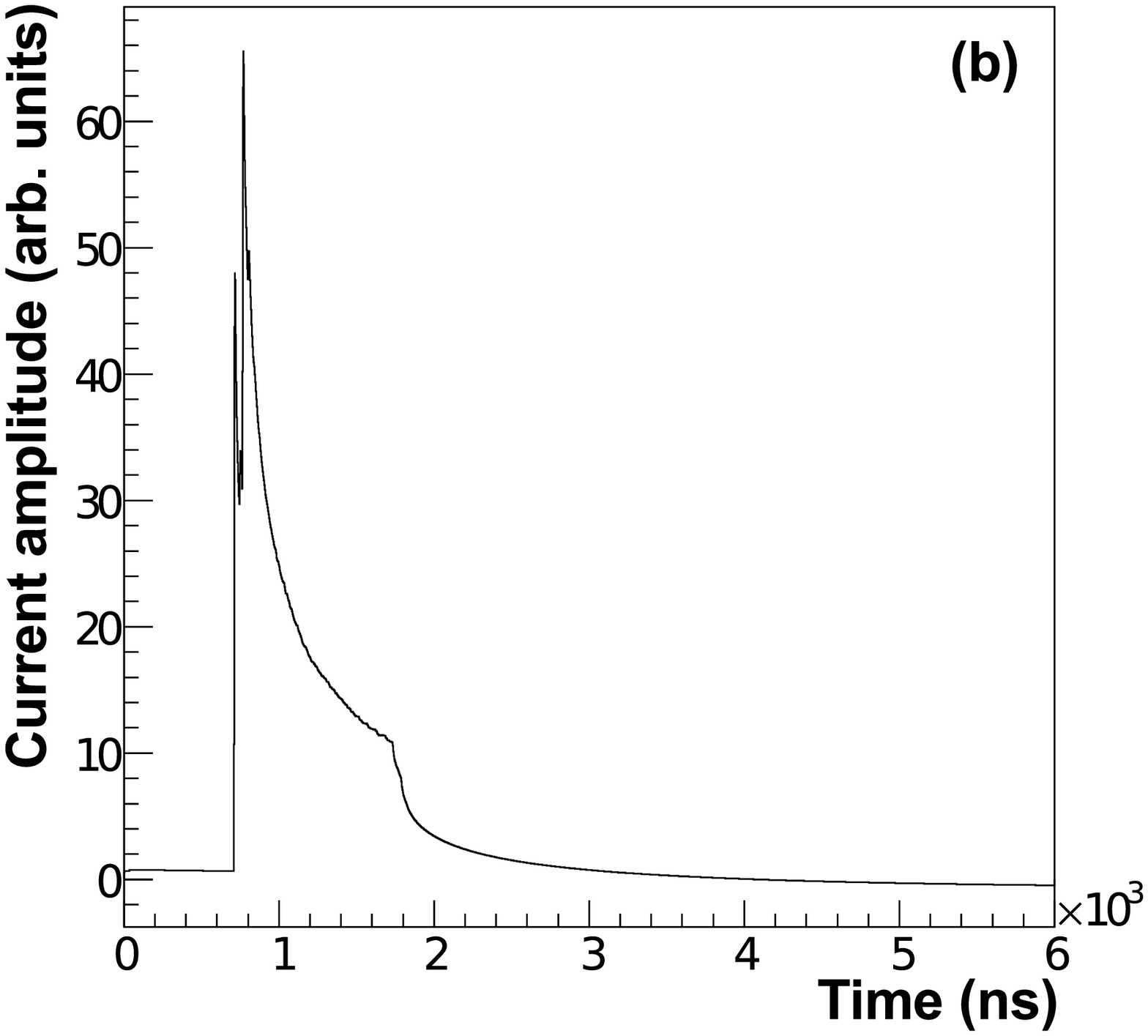}
\includegraphics[width=0.3\textwidth]{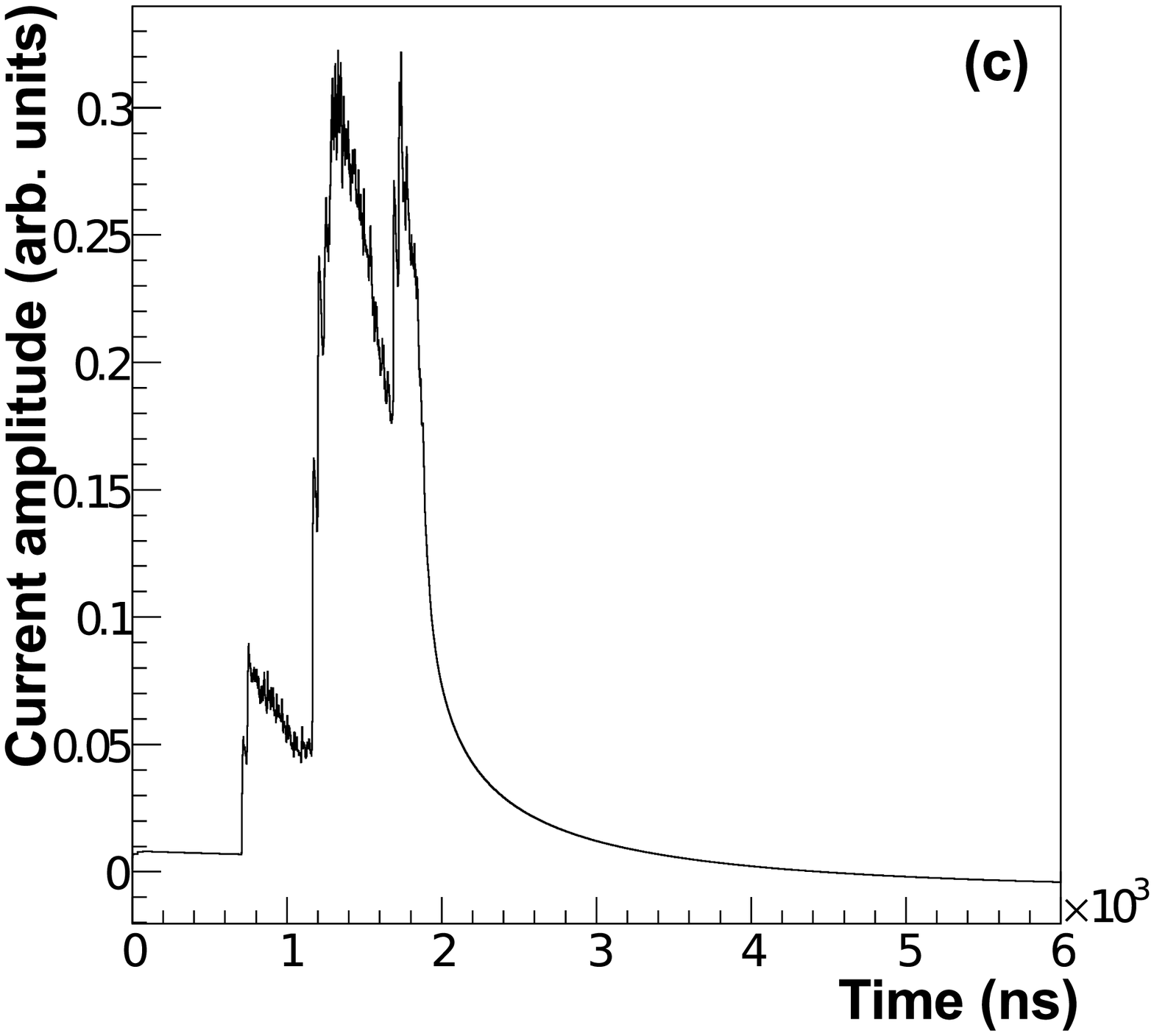} \\
\includegraphics[width=0.3\textwidth]{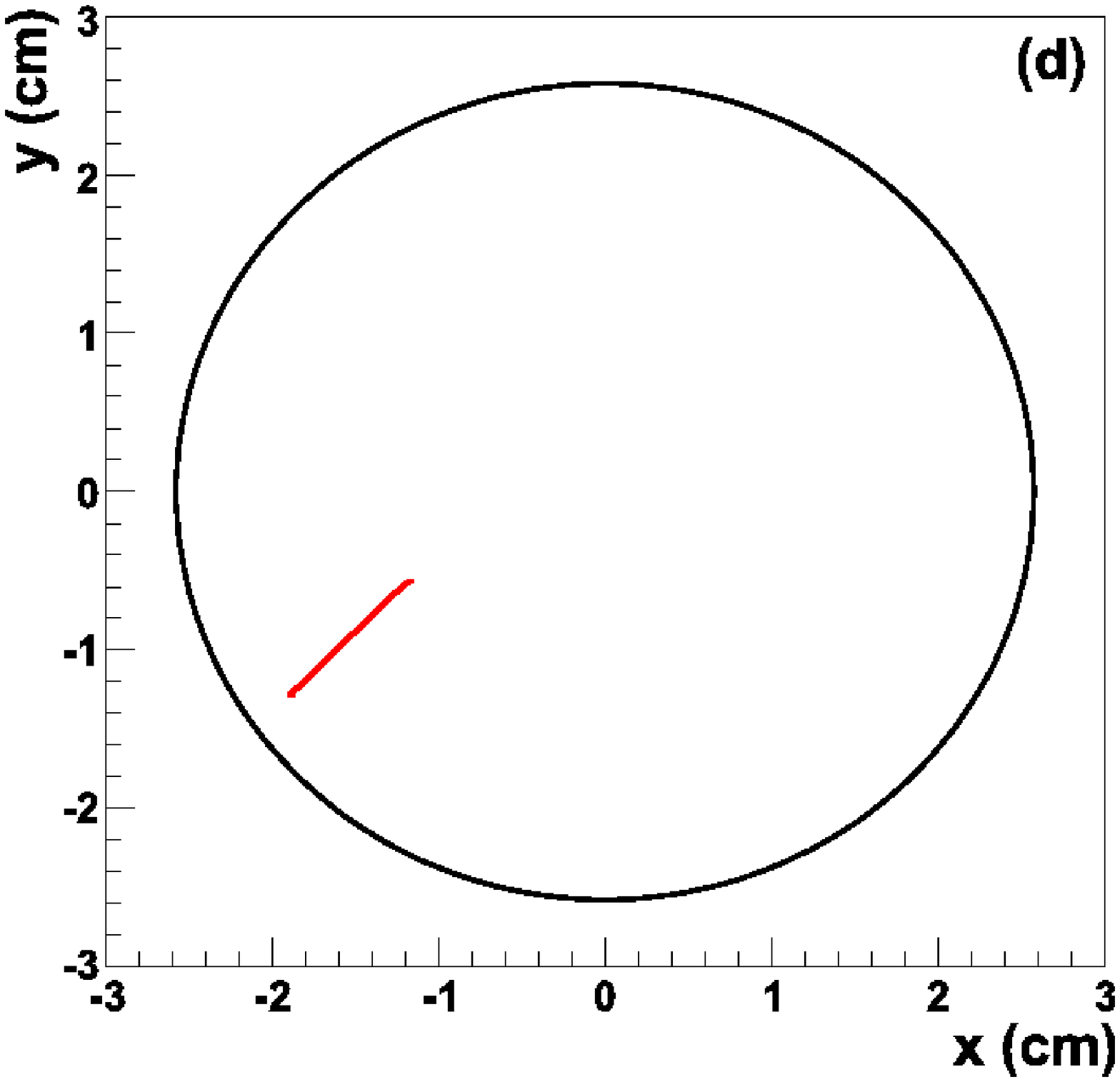}
\includegraphics[width=0.3\textwidth]{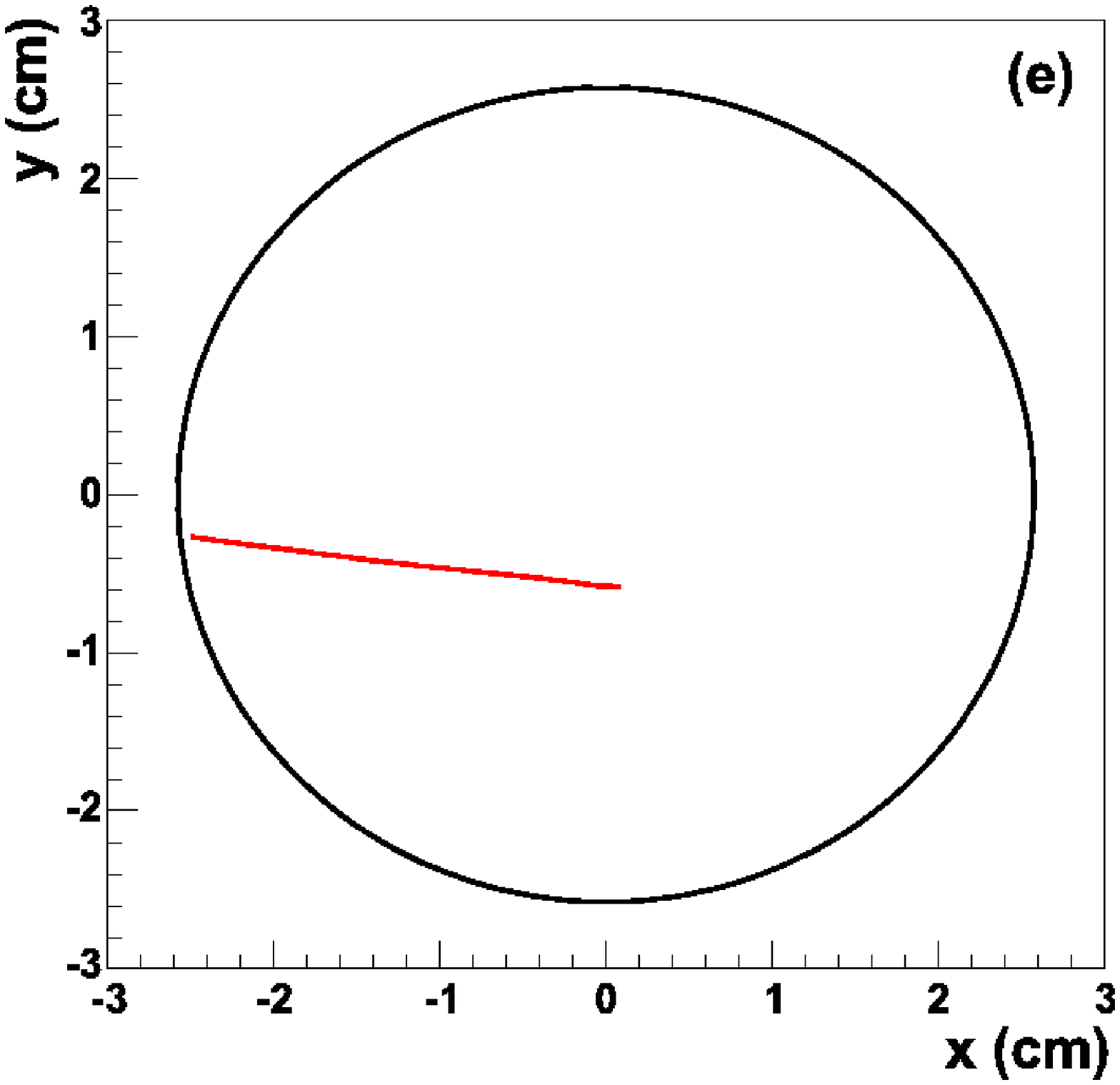}
\includegraphics[width=0.3\textwidth]{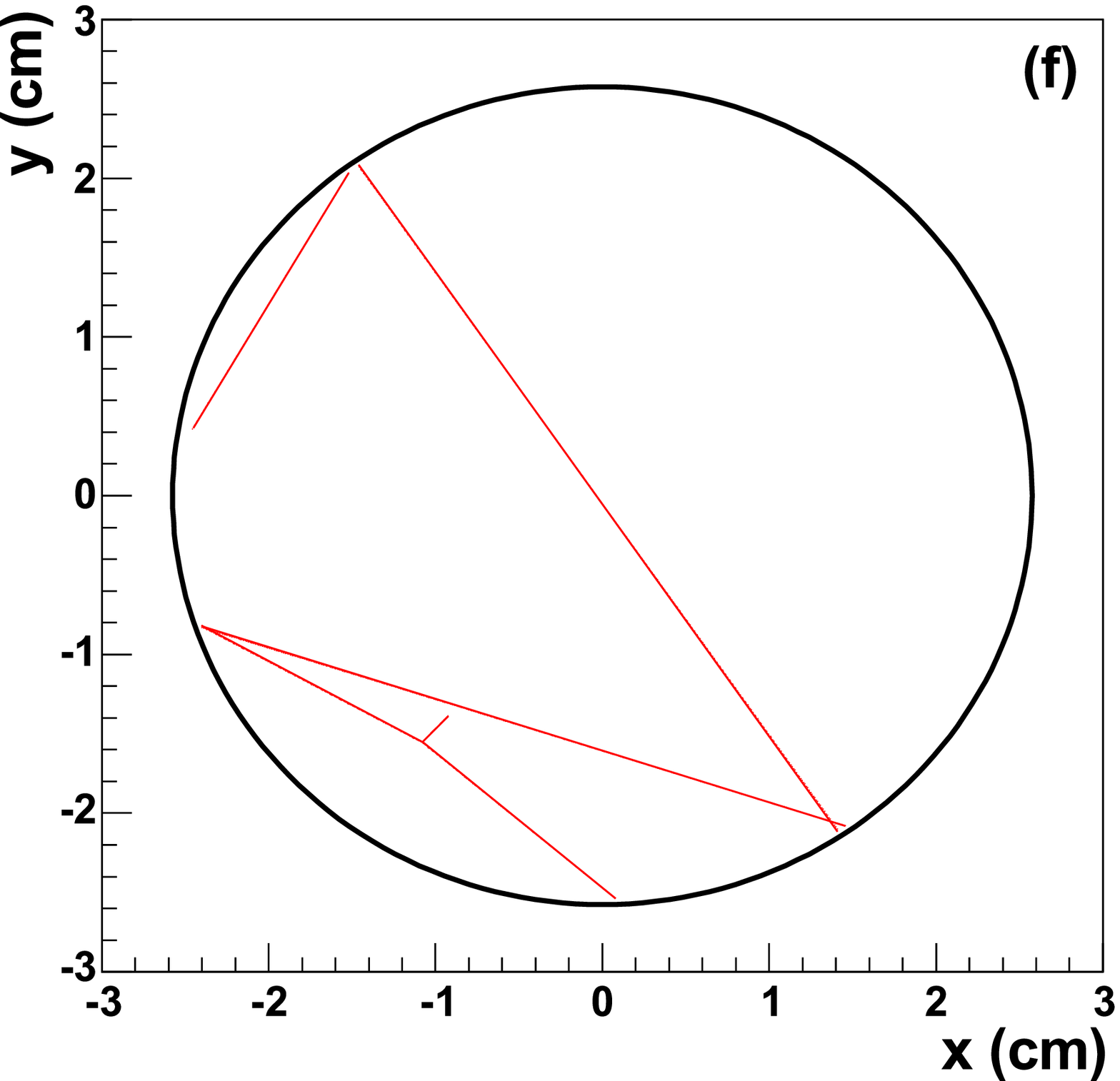}

\caption{\label{fig:examplepulses}
Representative pulses from the three classes of ionization events: (a) a neutron, (b) a 5~MeV alpha, and (c) a $\beta$ particle.  The particle trajectories viewed in the radial plane are shown in (d), (e), and (f), respectively. The track coordinates are given in the text.}
\end{figure}

The power of this numerical approach is that any pulse can be computed, given the location and number of ionization electrons in the event. Ignoring for the moment variations in the particle trajectory, it is convenient  to formulate pulses as a function of four variables, in addition to $t$: $I_{\mathrm{track}}(E,r,\theta,\phi)$, where $E$ is the particle energy, $r$ its starting radius, $\theta$ the track angle relative to the wire, and $\phi$ is an azimuthal angle relative to the radial vector (see \fref{fig:theta_and_phi}).  \Fref{fig:examplepulses} (top row) shows typical $I_{\mathrm{track}}$ shapes from the three classes of physics events of interest: a neutron capturing at $r=1.29$~cm, with the proton traveling outwards at $\theta=67^{\circ}$ and $\phi=13^{\circ}$ (first column), a 5~MeV alpha particle starting at $r=2.5$~cm, with $\theta=105^{\circ}$ and $\phi=14^{\circ}$ (second column), and a 5~MeV electron starting at $r=2.5$~cm and scattering thrice on the walls before being absorbed in the nickel. A small $\delta$-ray resulting from a M\"{o}ller scatter can be seen at $(x,y)=(-1.2,1.5)$~cm. Track projections onto the radial plane are displayed in the bottom row. 

\subsection{Ionizing Particles in the NCD Counters}\label{section:ioniz}
\subsubsection{Protons, Tritons and Alphas}\label{section:tracks}
In the energy range of interest (0.2-8~MeV), the energy loss, $dE/dx$, for protons, tritons and alphas in the NCD counter gas are accurately calculated by the software package TRIM~\cite{ref:TRIM}.\footnote{TRIM has been extensively verified against measurements of energy loss in He, C, F, and CF$_4$, with agreement at the few-percent level.} The mean stopping ranges of protons and tritons in the NCD counters are 0.73~cm and 0.28~cm respectively, corresponding to a mean electron arrival interval of $\sim$1000~ns for perpendicular, fully-contained proton-triton tracks. 

\begin{figure}
\begin{center}
\mbox{
\begin{minipage}{0.48\textwidth}
\begin{center}
\includegraphics[height=8.cm]{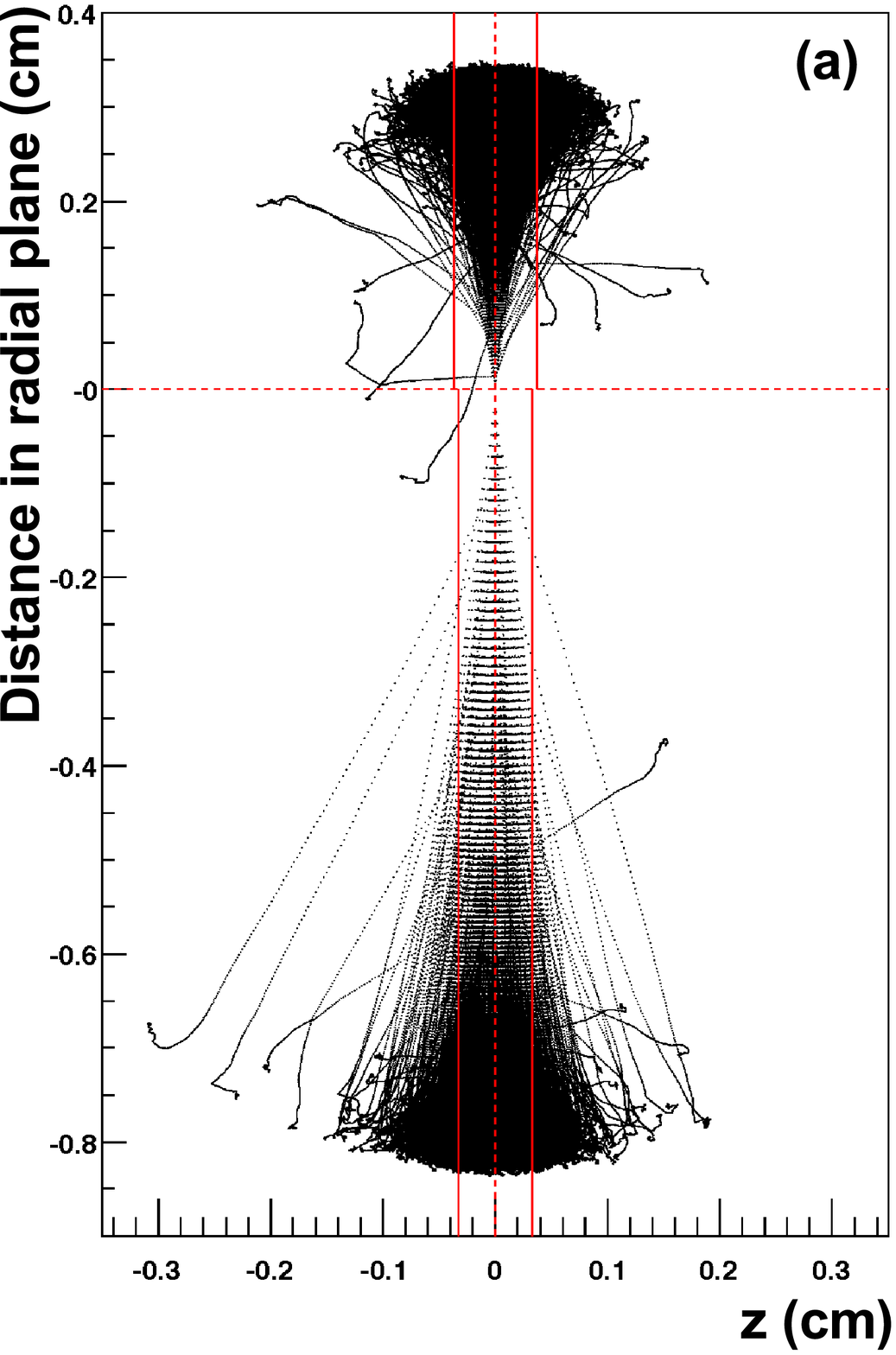}
\end{center}
\end{minipage}
\begin{minipage}{0.48\textwidth}
\begin{center}
\includegraphics[height=8.cm]{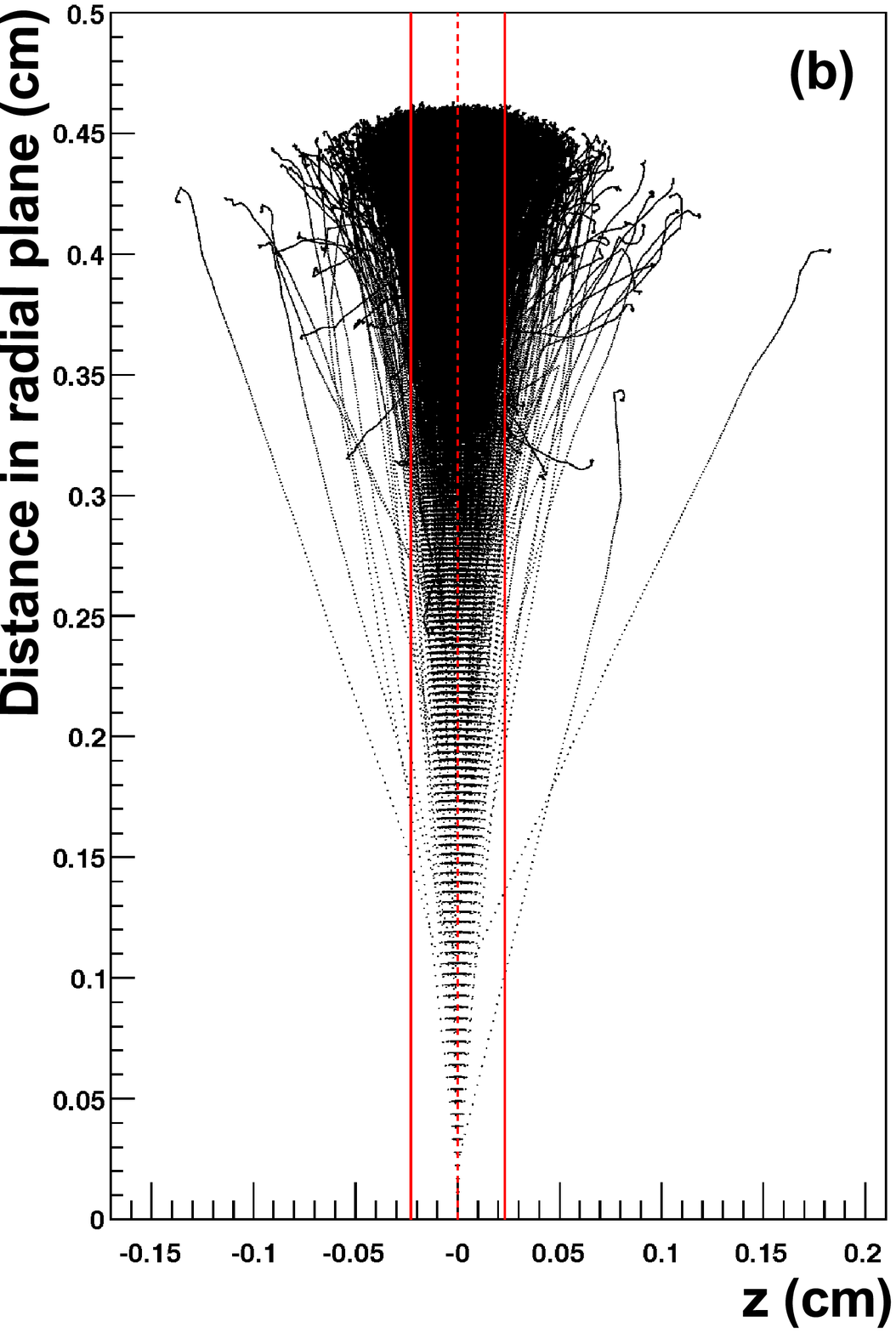}
\end{center}
\end{minipage}}
\caption{\label{fig:particletracks}
Simulation of particle trajectories in 85:15 $^3$He-CF$_4$, with the method described in~\sref{section:tracks}. All particles start at the origin; a conical shape results from early scatters in particle tracks. Left: $p$-$t$ tracks, with tritons directed upwards, and protons downwards. Right: 1~MeV $\alpha$ tracks. 1-$\sigma$ deviations of the track endpoints from straight-line travel are shown in solid red lines.}
\end{center}
\end{figure}

To take the effects of lateral straggling on time resolution into account, multiple scattering in the gas has to be considered. Particle trajectories in nickel and counter gas are calculated with the Ziegler-Biersack-Littmark method as described in~\cite{ref:ZBL}. \Fref{fig:particletracks} shows 3000 calculated {\it p-t} (left) and 1~MeV alpha tracks (center) in the NCD counter gas volume, all starting off in the same initial direction (parallel to the vertical axis) from the origin. The mean position of track endpoints on the horizontal axis is 0~cm, with RMS values of 0.33~mm, 0.37~mm and  0.23~mm for 573~keV protons, 191~keV tritons, and 1~MeV alphas, respectively.  The RMS of the distribution of track endpoints from our calculation is in good agreement with full TRIM Monte Carlo calculations~\cite{ref:TRIM}, differing by up to 16\% at 9~MeV.  More details on the simulation of ionizing particles in the NCD counters can be found in~\cite{ref:hwctthesis}.

\begin{figure}
\begin{center}
\includegraphics[height=8.cm]{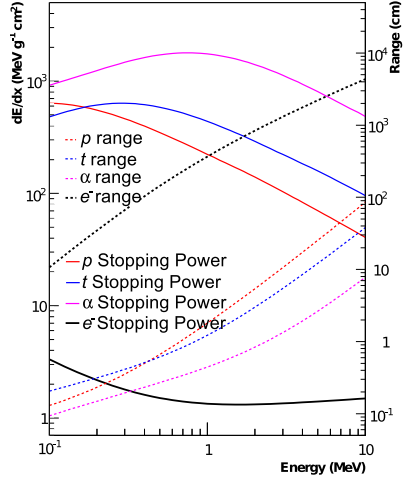}
\caption{\label{fig:stoppingranges}
$dE/dx$ (solid lines) and stopping ranges (dashed lines) of protons (red), tritons (blue), alphas (magenta) and electrons (black) in the NCD counter gas volume.}
\end{center}
\end{figure}

All tracks are assumed to appear instantaneously in the gas. This is a reasonable approximation, since a non-relativistic calculation yields flight times of $\sim$2~ns for the proton and $\sim$4~ns for the triton. Hence, any broadening effects on pulse shapes due to transit time of the primary particle are negligible. It is further assumed that all primary and secondary ion-pairs are created along the trajectory. According to Rudd {\it et al.}~\cite{ref:Rudd}, the kinetic energy imparted to ionized electrons in helium by $\sim$1~MeV protons is likely to be less than 1~keV. The range of 1~keV electrons in the NCD counter gas is less than 60~$\mu$m, or at most, only 0.5\% of the total length of a fully-contained $p$-$t$ track. This means that secondary ion pairs ({\it e.g.} occurring from ionization of CF$_4$ by ionization electrons) are created close to the particle path. The effect on pulse shapes, apart from a minor smearing effect, is negligible. 

\subsubsection{$\beta$ Particles}
The propagation of $\beta$ particles and $\gamma$s in NCD counter gas is handled by the software package EGS4~\cite{ref:EGS4}. The interactions of electrons with the NCD counter gas can be classified in three main categories: (1) inelastic collisions, including excitations and ionizations, (2) elastic interactions with electrons ({\it i.e.} M\"{o}ller scattering), and (3) Bremsstrahlung. The primary mode of energy loss is via inelastic collisions with gas particles, which can be evaluated with the Bethe-Bloch formula. The resulting energy loss per unit length, $dE/dx$, and range are shown in \fref{fig:stoppingranges}, as a function of energy. A typical $\beta$ track contains multiple elastic scatters on the nickel walls. This widens and complicates the structure of $I_{\mathrm{track}}$. On account of the very low stopping power, most $\beta$ pulses are low-amplitude and do not trigger the proportional counter. The probability of a triggered event resulting from an electron possessing more than 0.2~MeV of kinetic energy is $< 0.1\%$.  The estimated number of detected $\beta$ events above 0.2~MeV in SNO's third phase resulting from U and Th impurities was $\sim$1.3.

\subsection{Electron Drift Times}\label{section:drift}
A Monte Carlo-based low-energy electron transport simulation was developed to evaluate the mean drift times, $t_{\mathrm{d}}$, of electrons in the NCD counter gas mixture as a function of initial radius.  This simulation was used instead of other available simulations (such as GARFIELD~\cite{ref:GARFIELD}) because the latter were not well suited for situations where the electric field varies rapidly (e.g. near the anode wire), or where elastic scattering interactions are dominant over a wide energy range, such that the electron takes a relatively long time to equilibrate with the gas.

\begin{figure}
 \centering
\includegraphics[width=0.8\textwidth]{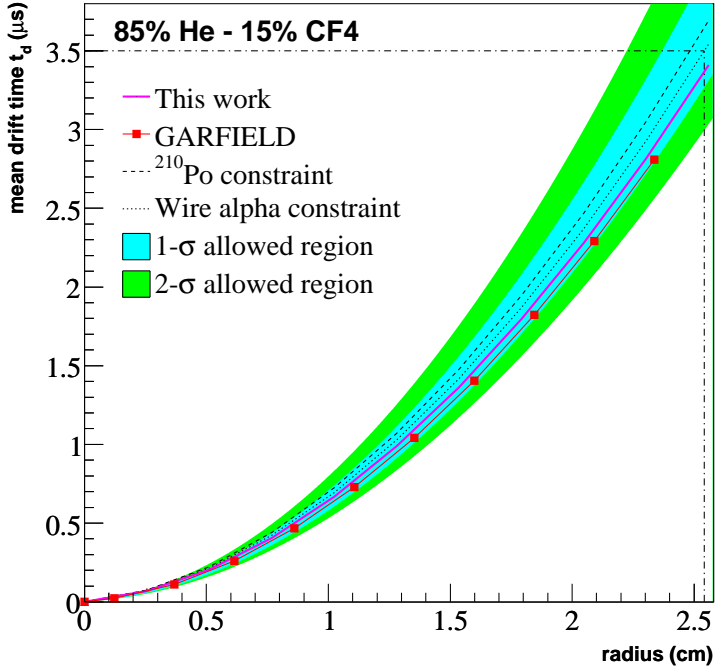}
\caption{\label{fig:rtcurve}
Mean electron drift time in the NCD counters as a function of radius. The cyan/light shaded region is the set of possible $t_{\mathrm{d}}(r)$, conservatively assuming existing measurements~\cite{ref:Kopp} to have an uncertainty of $\pm$10\% (none were given in~\cite{ref:Kopp}). The green/dark and cyan/light shaded regions combined represent a $\pm$20\% (2-$\sigma$) uncertainty. Regions above the dotted curve are disfavoured by wire $\alpha$ pulses, which require $t_{\mathrm{d}}(r=2.54 \mathrm{cm})<3451$~ns (denoted by the dashed-dotted lines). The dashed curve is a weaker constraint from low-energy $^{210}$Po events, while the magenta solid curve is the actual function adopted in pulse simulations. GARFIELD calculations are the red diamond data points.}
\end{figure}

The electron-transport simulation we developed showed good agreement with GARFIELD predictions in benchmarks using a constant electric field.  We also compared with past measurements by Kopp {\it et al}~\cite{ref:Kopp}. \Fref{fig:rtcurve} shows the 1-$\sigma$ (cyan/light shading) and 2-$\sigma$ (green/dark shading) allowed regions for $t_{\mathrm{d}}$ in a 85:15 $^3$He-CF$_4$ mixture, conservatively assuming experimental uncertainties of $\pm10$\% at all electric field values (no uncertainties are given in~\cite{ref:Kopp}). 
 
Further verifications can be made by inspecting specific types of alpha pulses. These include 5.3-MeV $^{210}{\rm Po}$ alphas that deposit between 0.9-1.2~MeV in the gas. The maximum radial length of these events is, on average, 0.12~cm, corresponding to an observed pulse FWHM of $\sim$400~ns. This means that, if one allows for a broadening of $\sim$100~ns by the electronic reflection, the time difference between drift times of electrons starting at 2.54~cm and 2.42~cm cannot exceed 300~ns. The dashed curve in \fref{fig:rtcurve} shows this requirement. A slightly more stringent constraint is obtained by examining the widest alpha particle events, which originate from the anode (dotted line in \fref{fig:rtcurve}). The drift times were increased by a scaling factor of 10$\pm4$\% to match the calculations to observed pulse-width distributions. Taking all of the above into account, the drift time curve for the NCD counter simulation, as a function of radius, is:

\begin{equation}\label{equation:driftcurve} 
t_{\mathrm{d}}=121.3r+493.9r^2-36.71r^3+3.898r^4
\end{equation}
with $t_{\mathrm{d}}$ in ns and $r$ in cm.  The uncertainty on $t_{\mathrm{d}}(r)$ was conservatively assumed to be $\pm$10\%, based on the agreement of the data and simulations shown in \fref{fig:rtcurve}.

Electron diffusion results in a radius-dependent smearing on all pulses, and dominates the time resolution. A smearing factor $\sigma_{\mathrm{D}}$ is determined as a function of $r$ from the electron-drift simulation, and is applied in pulse calculations. $\sigma_{\mathrm{D}}$ and $t_{\mathrm{d}}$ are found to be linearly related:
\begin{equation}
\sigma_{\mathrm{D}}(t_{\mathrm{d}})=0.0124 t_{\mathrm{d}} + 0.559
\end{equation}
The differences between the drift curves and the electron diffusion for $^{3}$He gas and $^{4}$He gas were found to be insignificant compared to the uncertainties in the model.

In the endcap regions of the NCD counters the electric field is not purely cylindrical (i.e. radial in the $x-y$ plane and constant along the $z$ axis).  Electrons were not propagated in the non-cylindrical-field region, since we do not know how the drift speeds or gas gain are affected.  The line determining approximately where the field transitioned from cylindrical to non-cylindrical was determined with a rough calculation of the fields in the endcap region.  Electrons from particles in the non-cylindrical-field region were not propagated.  We estimate that approximately 1\% of all neutron and alpha pulses were affected by the endcap region. The impact of these pulses in the data analysis is discussed in \sref{alphabackgrounds}.

\subsection{The Gas Gain}\label{section:gain}
The mean NCD counter gas-gain (averaging over many events in a counter), $\overline{G}$, as a function of voltage, is well described by the Diethorn formula~\cite{ref:Knoll}:
\begin{equation}\label{eq:diethorn}
\ln \overline{G} = \frac{V}{\ln(b/a)}\frac{\ln{}2}{W}\left(\ln\frac{V}{pa\ln(b/a)}-\ln{}K\right).
\end{equation}
The Diethorn parameters are the average electric potential change between ionization events, $W$, and the cylindrical electric field,
\begin{equation}
{\cal E}(r=r_{\mathrm{av}}) = \frac{V}{\mathrm{ln}(b/a)r_{\mathrm{av}}}.
\end{equation}
${\cal E}$ depends on the anode voltage, $V$, the NCD counter's anode-wire and inner-wall radii, $a$ and $b$ respectively, and the mean avalanche radius, $r_{\mathrm{av}}$.  All of these parameters were measured for the NCD counters and are listed in \tref{tab:scfixedparameters}.

However, the $\overline{G}$ can vary in several ways from the array average.  These changes can affect the detected energies and pulse shapes.  The gain on a particular string can be different from the array average if, for example, $V$ is slightly different on that string.  Within a string, $\overline{G}$ can vary by counter because of subtle differences in the gas pressure or inner wall radius, $b$.  The spread in gains between the counters was approximately 3\%.  String and counter variations in the mean gas gain were measured with neutron calibrations, and those variations were implemented in the NCD simulation.

Besides variations in $\overline{G}$, during the formation of a current pulse $G_i$ will differ from $\overline{G}$ for each track segment, $i$. There are random statistical fluctuations in the size of an ionization avalanche from a single electron.  For a mean gas gain of approximately 220, the avalanche size varies exponentially~\cite{ref:Kopp}.  These fluctuations are easily simulated by randomly choosing the gas gain from an exponential distribution with mean $\overline{G}$.  This effect results in a subtle smoothing of the pulse shape.

More significantly from the point of view of energy spectra and pulse shapes, under the typical NCD array operating conditions (i.e. for anode voltage $V = 1950$~V) the charge multiplication is sufficiently high for ion shielding to become non-negligible.  The energy spectra and wide-angle (large $\theta$) pulse shapes are substantially modified by this so-called ``space-charge'' effect.

A two-parameter model that accounts quantitatively for the space-charge effect was implemented. 
Consider a cluster of ions of total charge, $q$, formed in an electron cascade close to the wire, located at a mean radius $\bar r$. The charge induced by these ions on the anode modifies the local wire charge density.\footnote{In the steady state, the anode wire has a global charge density that depends on the applied voltage.} The change in gas gain, $\delta G$, resulting from a change in wire charge density, $\delta\lambda(\bar r)$, is derived from \eref{eq:diethorn}:
\begin{equation}\label{equation:dG}
\delta G \propto \overline{G}\mathrm{ln}\overline{G}\frac{\mathrm{ln}(b/a)}{2\pi\epsilon_0 V}
 \left(1+\frac{1}{\mathrm{ln}(r_{\mathrm{av}}/a)}\right)\delta\lambda(\bar r).
\end{equation}
$\delta\lambda$ is obtained by dividing the induced charge by a characteristic shower width in the spatial dimension parallel to the anode wire, ${\cal W}_s$, which, for simplicity, is assumed to be a constant for all avalanches:
\begin{equation}
\delta\lambda(r) = \frac{q}{{\cal W}_s} \frac{\mathrm{ln}(b/\bar r)}{\mathrm{ln}(b/a)}.
\end{equation} 
Electrons originating from a given segment, $i$, of a particle track are affected by the density changes brought about by ions formed in previous electron cascades, $\delta \lambda_j$. Each of these ion clusters moves slowly towards the cathode while the primary 
electrons are being collected. In the presence of many ion clusters, the total change in the anode charge density at time $t$, affecting the evolution of the electrons from the $i^{th}$ track segment, is therefore:			
\begin{equation} \label{equation:dlamb}
\delta\lambda_i = \frac{e}{{\cal W}_s}\sum_{j=1}^{i-1}
    \frac{\mathrm{ln}[b/\bar{r}_j(t)]}{\mathrm{ln}(b/a)}G_jn_{\mathrm{pair},j} + 
    \frac{e}{{\cal W}_s}\frac{\mathrm{ln}(b/\bar{r})}{\mathrm{ln}(b/a)}n_{\mathrm{pair},i}.
\end{equation}
$n_{\mathrm{pair},j}$ is the number of ion pairs formed in the $j^{th}$ segment, and $j$ loops over all the previous ion clusters, which have moved to different radii $\bar r_j(t)$ at time $t$. $\bar r_j(t)$ is solved by integrating the relation
\begin{equation}
\frac{\mathrm{d}r_j}{\mathrm{d}t}=\mu{\cal E}(r_j) \longrightarrow
 \bar r_j(t)^2=\frac{2\mu{}Vt}{\mathrm{ln}(b/a)}+r_{\mathrm{av}}^2,
\end{equation}
with ${\cal E}(r_j)$ being the value of the electric field at $r_j$, and $\mu$ the ion mobility.

A charge segment cannot have a significant impact on the gain of another segment if their avalanches are far apart in the $z$ direction. For simplicity the shower width is assumed to be a step function; an electron shower centered at a position $z_0$ on the wire is only affected by segments collecting within the limits $z_0-{{\cal W}_s} <z < z_0+{{\cal W}_s}$. This approximation is crude, but efficient.  For those which do overlap, the common distance  between cascades is calculated and the induced charge density weighted by an overlap factor $\xi$. As an example, a group of electrons arriving at the anode at $z_1<z_0$, with $z_1+{{\cal W}_s}/2>z_0-{{\cal W}_s}/2$ has an overlap factor of $({{\cal W}_s}+z_1-z_0)/{{\cal W}_s}$. \Eref{equation:dlamb} then becomes:
\begin{equation} \label{equation:overlap} \delta\lambda_i = \frac{e}{{\cal W}_s}\sum_{j=1}^{i-1}
    \frac{\mathrm{ln}[b/\bar{r}_j(t)]}{\mathrm{ln}(b/a)}G_jn_{\mathrm{pair},j}\xi_j + 
    \frac{e}{{\cal W}_s}\frac{\mathrm{ln}(b/\bar{r})}{\mathrm{ln}{b/a}}n_{\mathrm{pair},i}.
\end{equation}
It is implicitly assumed that ions produced in avalanches induce an image charge of uniform density along the wire.

The mean gas gain of the $i^{th}$ track segment is, therefore:
\begin{equation}
\overline{G}_i=\overline{G}-\delta G_i.
\end{equation}
$\delta G_i$ is evaluated with \eref{equation:dG}, using \eref{equation:overlap} as an input.

In this numerical model, the two parameters that need to be optimized are: [1] the constant of proportionality in \eref{equation:dG} (referred to as $\eta$ in \tref{tab:scfixedparameters}), and [2] the avalanche width ${\cal W}_s$. These two quantities share a strong inverse correlation. Their values are determined by tuning the $^{210}$Po peak position relative to the neutron peak, the position of the bump in the $^{210}$Po spectrum caused by the space charge relative to the $^{210}$Po peak, and the shape of the neutron spectrum.  The parameters ${\cal W}_s$ and $\eta$ are needed to accurately reproduce both neutrons and $^{210}$Po alphas, which implies that at least one of the two parameters varies with energy.  In this model all of the energy dependence is given to ${\cal W}_s$, while $\eta$ is fixed.  Further information about the electron-drift and gas-gain simulation can be found in~\cite{ref:hwctthesis}.

Other physics parameters, such as  $\overline{G}$, $\mu_i$, $W$ and $r_{\mathrm{av}}$, are constrained independently.  They are listed in \tref{tab:scfixedparameters}. The following two sections describe the measurements made to determine $W$ (\sref{section:W}) and $\mu$ (\sref{section:imob}).

\begin{table}[htb]
\caption{\label{tab:scfixedparameters}Space-charge model input values.  For ${\cal W}_s$, the energy $E$ is in MeV, and the uncertainties for both the gradient and offset are given.}
\begin{indented}
\lineup
\item[]\begin{tabular}{@{}lll}
\br
Parameter & Value & Uncertainty \\
\mr
$\eta$                            & $\0\0{}1.5$     & $\0{}0.1$   \\
${\cal W}_s$ ($\mu$m)             & $154E + 782$    & $31$, $120$ \\
$\overline{G}$                    & $219$           & $10$        \\
$\mu$ ($10^{-8}$ cm$^2$/ns/V)     & $\0\0{}1.082$   & $\0{}0.027$ \\
$W$ (eV)                          & $\0{}34$        & $\0{}5$     \\
$r_{\mathrm{av}}$ ($\mu$m)        & $\0{}58$        & $10$        \\
\br
\end{tabular}
\end{indented}
\end{table}

\subsection{The Gas Gain and the Average Energy per Ion Pair}\label{section:W}
Two parameters are primarily responsible for determining the integrated current, $I$, measured with a proportional counter in response to a given mean amount of energy, $\overline{E}$, deposited in the gas: the average energy deposited per ion pair created, $W$, and the gas gain, $\overline{G}$.  Integrating over many current pulses, the relationship between $E$, $W$, and $\overline{G}$ is~\cite{ref:Knoll}
\begin{equation}
\frac{W}{\overline{G}} = \frac{n e \overline{E}}{I},
\end{equation}
where $n$ is the rate of neutron captures and $e$ is the  electron charge.  $W$ is a characteristic of the ion and the medium in which it is traveling.  In this simulation we have assumed that $W$ is the same for protons, tritons, and alphas; and that $W$ is independent of energy. This is valid except at low energies, which will have little effect on the NCD pulses~\cite{ref:icrureport31}.

To our knowledge $W$ has not been measured before for the gas mixture used in the NCD counters, but the values for protons and alphas in a variety of other gases are all fairly similar to each other~\cite{ref:icrureport31}. To obtain a precise value of $W$ for the SNO counters, we have measured $\overline{G}/W$ both in the ion saturation region (low voltage), and at the operating voltage of the NCD counters.

We conducted this test with an undeployed NCD counter. Three radioactive sources were used simultaneously to provide a large neutron flux: $^{241}$AmBe, $^{252}$Cf, and Pu-$^{13}$C. They were set within approximately 30~cm of the NCD counter with blocks of polypropylene in between to act as a neutron moderator. A layer of aluminum foil was wrapped around the counter as a Faraday cage to avoid currents induced on the NCD counter by external electromagnetic fields. The foil was separated from the NCD counter body by a layer of plastic bubble wrap, and it was electrically connected to ground. The counter was first set up with standard SNO data acquisition hardware to determine the event rate which, after being corrected for deadtimes, was $(429.9 \pm 1.1) \,\mathrm{Hz}$.

The second setup replaced the SNO data acquisition system with a picoammeter to read the DC current from the NCD counter.  Instead of reading the current from the digital display on the picoammeter, which fluctuated rapidly due to statistical variations at the low current levels we were measuring, the analog output was fed into a digital oscilloscope that averaged the reading over periods of a few seconds. It was found that the average value would stabilize reliably within that amount of time. The high-voltage supply was connected to the anode wire, and its setting was controlled by the data acquisition system. The picoammeter read the current from the cathode (i.e. the nickel wall of the NCD counter).

\begin{figure}[htbp]
\begin{center}
\includegraphics[width=5in]{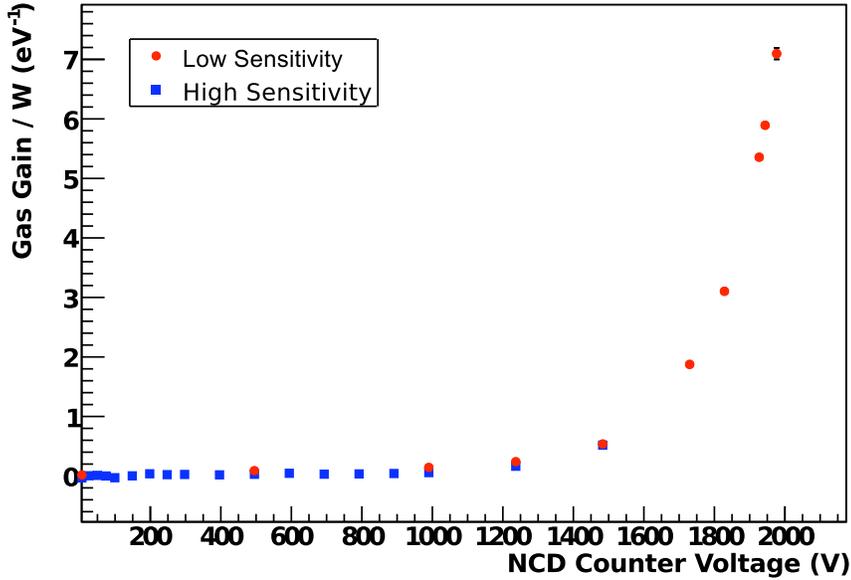}
\caption{The two measurements of $\overline{G}/W$.  The uncertainties plotted are statistical only, and are smaller than the data-point markers for almost every point.  The square points are a high-sensitivity measurement using the lowest measurement setting on the picoammeter.  The circle points are a lower-sensitivity measurement that goes to higher voltages, using the next-higher picoammeter setting.  The both measurements agree well, though there may be a small systematic shift between the two of them that is accounted for by the systematic error due to the picoammeter accuracy.}
\label{fig:highvmeas}
\end{center}
\end{figure}

The results from these measurements are shown in \fref{fig:highvmeas}. The lowest measurement setting on the picoammeter (range: 2~nA) was able to make a high-sensitivity measurement of the current up to 1500~V, which is shown with the squares in \fref{fig:highvmeas}.  The next-higher current range (20~na) was used to make a lower-sensitivity measurement up to 2000~V (circles in \fref{fig:highvmeas}).  Due to differences in how the high-voltage supply was calibrated relative to the supplies used in the actual NCD array system, the setting which corresponded to the voltage used for the NCD array was 1943.9~V.  The data point at this voltage determines $\overline{G}/W$ to be:
\begin{equation}\label{eq:goverwresult1}
\overline{G}/W = 6.36 \pm 0.33 \mathrm{(stat)} \pm 0.03 \mathrm{(syst)} \mathrm{eV}^{-1}.
\end{equation}
Using the value of $\overline{G}=219$ as shown in \tref{tab:scfixedparameters}, $W$ was found to be $34 \pm 5$~eV.

The high-sensitivity measurements can be used as a comparison.  A fit with a zeroth order polynomial in the ion saturation region, between 200 and 800~V, determines $\overline{G}/W$:
\begin{equation}\label{eq:goverwresult2}
\overline{G}/W = [2.93 \pm 0.65 \mathrm{(stat)} \pm 0.84 \mathrm{(syst)}] \times 10^{-2} \mathrm{eV}^{-1}.
\end{equation}
Since $G=1$ in the ion saturation region, we find $W = 34.13 \pm 12.4$~eV.  These results are in excellent agreement with each other, and the value of $W = 34 \pm 5$~eV was implemented in the NCD array MC.

\subsection{Ion Mobility}\label{section:imob}
The small ion mobility, relative to that of the electrons, results in the long tail that is characteristic of pulses from ionization in the NCD counters, as shown in \fref{fig:pulsetransform}. Therefore it is important to know the ion mobility so that the tail of each pulse can be simulated correctly.

The evolution of a current pulse in a cylindrical proportional counter is described in \eref{equation:wilkinson}.  $\tau$ is the ion time constant, which is inversely proportional to the ion mobility, $\mu$:
\begin{equation}
\tau = \frac{a^2 p \ln(b/a)}{2 \mu V_0};
\end{equation}
$V_0$ is the applied voltage, $p$ is the gas pressure, and $a$ and $b$ are the radii of the anode wire and the inner radius of the NCD counter, respectively. 

If the underlying shape of a pulse, not including the tail, is known, it is simple to extract the shape of that tail.  There is one class of pulses that has a relatively simple underlying structure: ionization tracks that are parallel to the anode wire.  All the primary ionization electrons reach the anode at approximately the same time, with some spread due to straggling.  Therefore the basic shape of the underlying pulse is approximately Gaussian.  Reflections and the effects of propagation along the counter and through the electronics also affect the shape of the pulse.  These secondary effects have all been modeled independent of the ion mobility.

\begin{figure}[htbp]
\begin{center}
\includegraphics[width=5in]{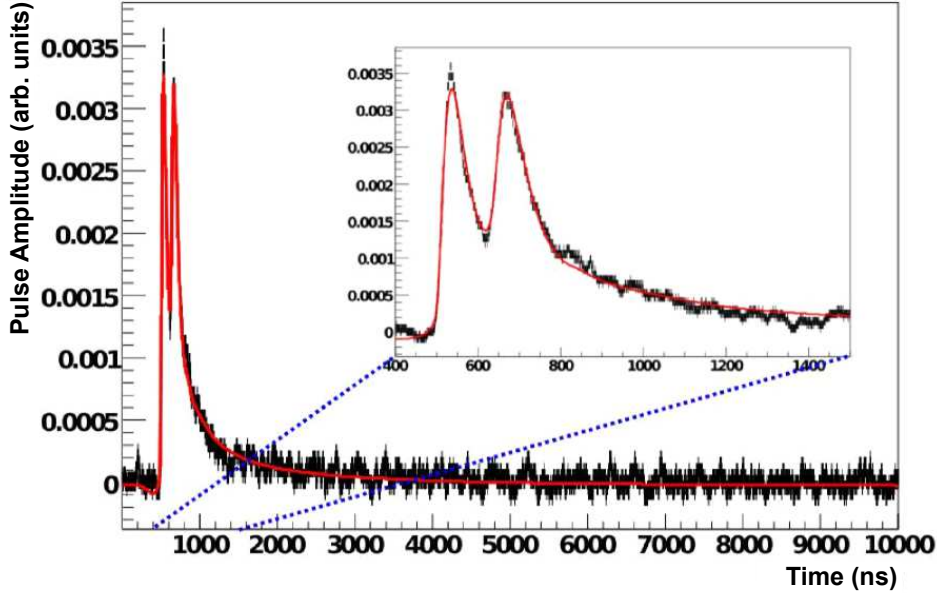}
\caption[Ion-tail time constant fit]{An example of a fit to extract the value of the ion-tail time constant.}
\label{fig:iontailfit}
\end{center}
\end{figure}

The procedure for extracting the ion mobility is to select the narrowest neutron pulses from a calibration data set and fit each pulse with a Gaussian convolved with a reflection and the electronics model.  The free parameters in each fit are $\tau$, the three Gaussian parameters (amplitude, mean and width), and the reflection time.  A fit example is shown in \fref{fig:iontailfit}.  The generic pulse model fits the peaks well enough to allow for a characterization of the ion tail.

A subset of the AmBe neutron calibration runs were analyzed in this study.  An initial selection of pulses was made by restricting the ADC charge to be between 100 and 150.  The neutron peak (764~keV) typically falls between 120 and 130~ADC~counts, so these are pulses where neither the proton nor the triton hit the wall.  A second selection cut was based on the width and height of each pulse.  The sharpness of a pulse can be approximately characterized by the ratio of the amplitude to the width.  A cut of $0.5\times10^{-4} < \mathrm{amplitude}/\mathrm{width}\ \mathrm{(V/ns)} < 1\times10^{-4}$ removed 99.36\% of the pulses.

\begin{figure}[htbp]
\begin{center}
\includegraphics[width=5in]{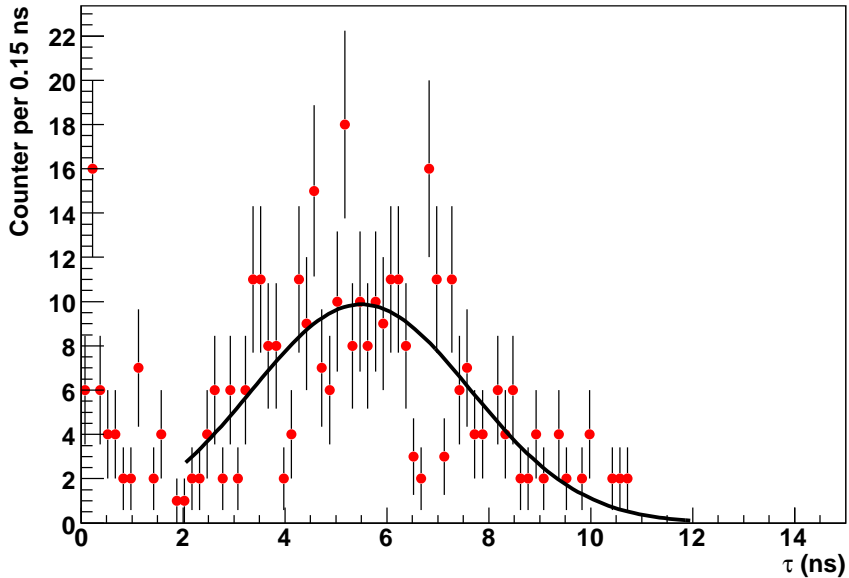}
\caption[Measurement of the ion-tail time constant]{Fit results for the ion-tail time constant.  The data set includes 393 pulses.  The low-$\tau$ peak is due to non-physics background pulses.  By fitting a Gaussian to the main peak with a log-likelihood minimization the ion time constant is determined to be $5.50 \pm  0.14$~ns.}
\label{fig:taufithistogram}
\end{center}
\end{figure}

\Fref{fig:taufithistogram} shows the results from all of the fits in the data set.  The histogram has a broad peak of successful fits, and a smaller peak at low $\tau$ of non-physics background pulses (e.g. spikes from electrical discharges do not have an ion tail).  Of the 393 pulses fit, 337 ($\approx$86\%) are above $\tau = 2$~ns.  The main peak was fit to a Gaussian using a log-likelihood minimization because of the small number of entries in many of the bins.  The mean of the Gaussian fit is $\tau = 5.50 \pm 0.14$~ns ($\sigma = 2.15 \pm 0.12$~ns).  The uncertainty is the statistical uncertainty of the fit for $\tau$ (the effects of changing the pulse-selection criteria were small compared to this uncertainty).  This result is expected to be correlated with the time constants of the electronics, which were measured \emph{ex situ} (i.e. the RC constants discussed in \sref{section:electronics}).

The time constant corresponds to a mean ion mobility of $\mu = (1.082 \pm 0.027) \times 10^{-8}$~cm$^2$~ns$^{-1}$~V$^{-1}$.  This value of the ion mobility was implemented in the NCD array MC, and the uncertainty was used to calculate its systematic effects.

\section{Electronics and Data Acquisition Simulation}\label{section:electronics}
\subsection{Pulse Propagation in the Counters}\label{section:pulseprop}
\begin{figure}[htbp]
\begin{center}
  \includegraphics[width=0.8\textwidth]{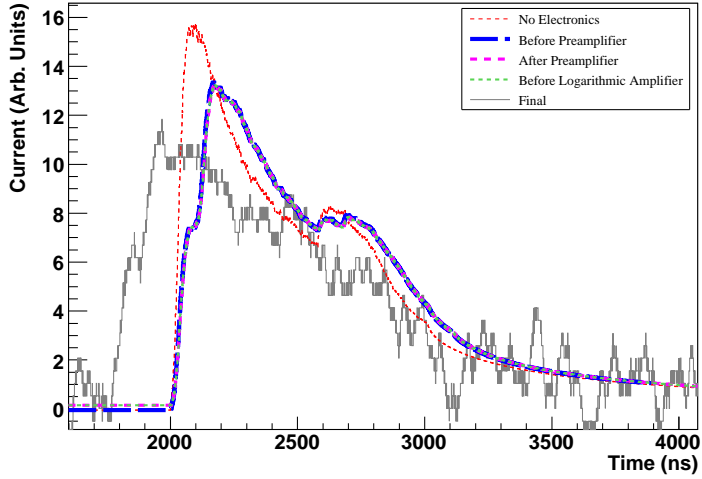}  
\caption{A simulated neutron pulse ($r = 1$~cm, $\theta = 90^{\circ}$, $\phi = 0^{\circ}$) at different stages of the electronics simulation to show the effects of the different components.  The most significant changes to the pulse shape are due to propagation in the NCD counters, and the logarithmic amplification ($\Delta t$ in \eref{eq:logamp}).  The shift of the start of the pulse is the time delay in the logarithmic amplifier.  The preamplifier, on the other hand, has little effect on the pulse shape.}
\label{fig:simstages}
\end{center}
\end{figure}

In the NCD array simulation, after a current pulse forms on an anode wire, it propagates along the counter through the NCD cable to the preamplifier. The amplified pulse is then transferred to the multiplexer system, at which point it is split between the two data-acquisition paths. One path integrates the pulse with a Shaper-ADC to determine the energy deposited in the counter. Its trigger is based on the pulse integral. The second path is triggered by the pulse amplitude. The pulse is logarithmically amplified and digitized with a sampling rate of 1~GHz. Each recorded pulse is 15~$\mu$s long. The electronic and data acquisition components are described in more detail in~\cite{ref:ncd_nim}.  Each current pulse is simulated using a 17,000-element array with 1-ns bin widths; a 15,000-element subset of the array is eventually stored in the standard SNO data structure for each pulse that causes a trigger. The extra 2,000 array elements are used to ensure that the beginning and end of each pulse are treated as if the current from the NCD counter exists continuously beyond the 15-$\mu$s-long pulse.  The start of the recorded 15-$\mu$s pulse is determined by the trigger conditions, downstream of the pulse simulation.  \Fref{fig:simstages} shows a simulated neutron pulse at various stages of the electronics simulation.

\subsection{Electronics simulation}

\begin{figure}[htb]
\begin{center}
\includegraphics[width=2in]{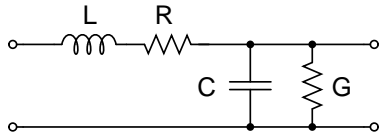} 
\caption{Circuit diagram for the Lossy Transmission Line model, showing the inductance (L), resistance (R), capacitance (C), and conductance (G).}
\label{fig:ltl}
\end{center}
\end{figure}

Propagation of the pulse along the NCD string is simulated with a Lossy Transmission Line (LTL) model~\cite{ref:lossytransline}. \Fref{fig:ltl} is a diagram of the LTL-model circuit.  Half of the pulse is propagated down the NCD string, through the delay line, and back to the point of origin of the pulse. The delay line attenuation is also simulated as a LTL. Both halves of the pulse (reflected and direct) are subsequently transmitted up to the top of the NCD string. The attenuation of the pulse due to transmission along the NCD string is dependent on the distance traveled, so pulses starting at different $z$ positions will look slightly different when they exit the string.

The simulation parameters for the NCD counter wire and delay line come from fitting the LTL model implemented  with the SPICE simulation package~\cite{ref:spice} to \emph{ex situ} electronics calibration data. The best fit parameters are given in \tref{tab:ltlparams}.  Skin effects in the anode wire, the resistances for the counters, and the delay lines result in frequency-dependent values for the parameters of the LTL model. To take the frequency dependence into account, the measured resistance ($R$, in $\Omega$/cm) is fit between 0 and 200~MHz with an empirical formula that illustrates the DC and frequency-dependent contributions:
\begin{equation}\label{eq:resistancemodel}
R(f) = \frac{A}{\exp[(f-B)/C] + 1} + \frac{D \sqrt{f} + E}{\exp[(B-f)/C]+1},
\end{equation}
where the fit parameters, $A$, $B$, $C$, $D$, and $E$ are given in \tref{tab:ltlparams}, and the frequency is given in MHz.

A LTL model using parameters calculated analytically, assuming an infinite cylindrical geometry, gave somewhat different results.  However, since the analytical model did not include the effects of counter endcaps, junctions, and the frequency-dependent permeability of the nickel walls, the SPICE fit was deemed superior, and those parameters were chosen for usage in the NCD array simulation.

\begin{table}[htdp]
\caption{The parameters used in the Lossy Transmission Line model of the NCD counters and of the delay line.  The resistance parameters are not all resistances because, as is shown in \protect\eref{eq:resistancemodel}, parameters B and C determine the frequency dependence of the different contributions.}
\begin{indented}
\lineup
\item[]\begin{tabular}{@{}ll}
\br
Parameter & Value \\
\mr
\multicolumn{2}{c}{Counter} \\
\mr
Inductance (10$^{-8}$ H/cm) & 1.33  \\
Capacitance (10$^{-14}$ F/cm) & 7.68  \\
Conductance (S/cm/MHz) & 0.0  \\
\mr
\multicolumn{2}{c}{Delay Line} \\
\mr
Inductance (10$^{-7}$ H/cm) & 9.91  \\
Capacitance (10$^{-12}$ F/cm) & 5.53  \\
Conductance (10$^{-12}$ S/cm/MHz) & 3.0  \\
\mr
\multicolumn{2}{c}{Counter and Delay Line} \\
\mr
Resistance - A ($\Omega$/cm) & $\0{}0.1024$  \\
Resistance - B (MHz) & $13.4$  \\
Resistance - C (MHz) & $\0{}8.4$  \\
Resistance - D (10$^{-2}$ $\Omega$/cm/$\sqrt{\mathrm{MHz}}$) & $\0{}1.643$  \\
Resistance - E (10$^{-2}$ $\Omega$/cm ) & $\0{}2.32$ \\
\br
\end{tabular}
\end{indented}
\label{tab:ltlparams}
\end{table}

Propagation in the NCD cable is simulated with a low-pass filter with an ${\rm RC}\approx 3$~ns. There is a small reflection of 15\% magnitude at the preamplifier input due to the slight impedance mismatch between the preamplifier input and the cable\footnote{The frequency-dependence of the reflection coefficient is not known, and it was assumed to be constant.  The 15\% reflection coefficient was found to match the data well}.  The fraction of the pulse that reflects off the preamplifier input travels to the bottom of the cable, reflects off the top of the NCD string, and subsequently travels back up the cable.

The preamplifier converts the current pulse to a voltage pulse, with a gain of 27500~V/A.  The circuit elements of the preamp also affect the shape of the pulse.  We simulate this with a low-pass filter (${\rm RC}\approx 22$~ns)  and a high-pass filter (RC~=~58,000~ns).  The RC constants were measured by fitting the model to \emph{ex-situ} injected pulses.

The implementation of the low- and high-pass filters performs the calculation in a single loop over the pulse array ($\propto N$, $N=17,000$~elements). Furthermore, in the case where the RC constant approaches the size of the bin width, the bins are subdivided to maintain the accuracy of the simulation. The low-pass filter is implemented as follows:
\begin{eqnarray}\label{eq:lowpassfilter}
\overline{V}_0 & = & \frac{d}{2\tau_{\mathrm{RC}}} V_0 \nonumber \\
\overline{V}_i & = & \left(\overline{V}_{i-1} + \frac{d}{2\tau_{\mathrm{RC}}} V_{i-1}\right) e^{d/\tau_{\mathrm{RC}}} + \frac{d}{2\tau_{\mathrm{RC}}} V_i, \qquad i \in [1, N)
\end{eqnarray}
where $V$ and $\overline{V}$ are the pulse before and after passing through the filter, respectively. $d$ is the bin width and $\tau_{\mathrm{RC}}$ is the RC time constant. The high-pass filter implementation is similar:
\begin{eqnarray}\label{eq:highpassfilter}
\overline{V}_0 & = & \left(1 - \frac{d}{2\tau_{\mathrm{RC}}}\right) V_0 \nonumber \\
\overline{V}_i & = & \left[\overline{V}_{i-1} - \left(1 + \frac{d}{2\tau_{\mathrm{RC}}}\right) V_{i-1}\right] e^{d/\tau_{\mathrm{RC}}} + \left(1 - \frac{d}{2\tau_{\mathrm{RC}}}\right) V_i
\end{eqnarray}

The multiplexer branch of the electronics chain consists primarily of a $\sim\,300$-ns delay cable, a logarithmic amplifier, and a digital oscilloscope.  The frequency response of the delay cable and the electronics between the preamplifier and the logarithmic amplifier are simulated with a low-pass filter (RC~$\approx$~13.5~ns). The analytic form of the logarithmic amplification is
\begin{equation}\label{eq:logamp}
V_{\mathrm{log}}(t) = a \log_{10}\left(1 + \frac{V_{\mathrm{lin}}(t-\Delta{}t)}{b} \right) + c_{\mathrm{chan}} + V_{\mathrm{PreTrig}},
\end{equation}
where $V_{\mathrm{log}}$ and $V_{\mathrm{lin}}$ are the logarithmic and linear voltages, respectively, and $a$, $b$, $c_{\mathrm{chan}}$, $\Delta{}t$, and $V_{\mathrm{PreTrig}}$ are constants determined by regular \emph{in situ} calibrations during data taking.  The circuit elements after the logarithmic amplification are simulated with the final low-pass filter (RC~$\approx$~16.7~ns).  The RC constants for the two low-pass filters in the multiplexer simulation were determined by fitting the model to pulses injected into the components \emph{ex situ}.  The final element of the multiplexer branch of the simulation is the digital oscilloscope.  The pulse array values are rounded off to the nearest integer to replicate the digitization.

\subsection{Noise simulation}

There are a variety of electronic noise sources within the NCD system.  Due to the difficulty of identifying and measuring all of the individual contributions, noise is added to the pulses after the rest of the simulation is complete. This situation also requires that noise is added to the multiplexer and shaper branches of the electronics independently.  For the multiplexer branch, the frequency spectrum of the noise was measured for each channel using the baseline portions of injected calibration pulses. This provides the mean value, $\mu_i$, of the noise power spectrum at each frequency. Assuming that the real and imaginary components of the noise are independent Gaussian-distributed random variables, then the standard deviation of the noise is related to the mean value by
\begin{equation}\label{eq:noise}
\mu_i = 2 \sigma_i^2.
\end{equation}
To avoid passing the noise through the final low-pass filter in the linear domain, the final simulated pulses without noise are ``de-logged'' (by inverting \eref{eq:logamp}), convolved with the noise, and subsequently ``re-logged.''

The Shaper branch of the electronics is simulated by a sliding-window integral of the preamplified pulse.  This number is then converted to units of ADC counts by doing an inverse energy calibration.  The constants used in the ``uncalibration'' are the same constants that are used to calibrate the data.  Since the shaper simulation acts on electronic-noise-free pulses, noise is added to the shaper value with a Gaussian-distributed random number.  The mean and standard deviation of the noise for each channel were determined by comparing the energy spectra from \emph{in-situ} neutron calibrations to the simulated electronic-noise-free energy spectra.  The noise characteristics of each NCD string were determined independently.  The typical RMS noise (in units of ADC values) is 2.0, with a variance of 0.7 (roughly $\mathrm{RMS}=0.012 \pm 0.004$~MeV) across the array.

\subsection{Trigger simulation}

Once all elements of each system are simulated for a given pulse, a trigger decision is made in the simulation based on whether the MUX or Shaper thresholds are exceeded. The MUX and Shaper systems include independent triggers and the thresholds for both are determined by \emph{in-situ} calibrations.  After a MUX trigger, the system is open for further triggers for 15~$\mu$s, after which all MUX channels are dead for 1~ms. The oscilloscope recording the pulse is dead for 0.9~s after it is recorded (the other oscilloscope is still live, provided it is not already reading out an event). The oscilloscope used to read out an event is determined by which is not busy, or by toggling between them if neither is busy.  After a shaper trigger, the system is open to further triggers for 180~ns, after which all shaper channels are dead for 350~$\mu$s. These times are only simulated within each Monte Carlo event and not between events. For instance, a single Monte Carlo event can involve the spontaneous fission of a $^{252}$Cf nucleus during a calibration.  Such a fission releases multiple neutrons and will result in multiple multiplexer and shaper events. The dead-time will be then simulated, but it will not apply between multiple $^{252}$Cf events.

All NCD-system triggers are then integrated with the SNO photomultiplier (PMT) signals.  The PMT trigger simulation time-orders an array of all PMT signals (i.e. each individual hit on a PMT) in a Monte Carlo event and scans through it to determine if any of the trigger conditions are satisfied (the trigger window is 430~ns long). If that occurs, then a ``global trigger'' is created and the simulated data is recorded. The NCD array signals (i.e. each pulse from the NCD system) are integrated with the PMT trigger simulation by inserting each NCD signal into the time-ordered array of PMT signals. As the simulation scans over the combined PMT+NCD signals, any individual NCD signal is sufficient to cause a global trigger.

\subsection{Shaper-Only Simulation}
Certain parts of the full NCD simulation described above are relatively slow due to the number of calculations that must be made.  For instance, the space-charge simulation requires a nested loop over the segments of an ionization track, performing calculations of the effects on the gas gain $\sim\,N^2$ times, where $N=17,000$. The electronics and data acquisition simulation also include several $\propto\,N$ and $\propto\,N\sqrt{N}$ loops over the pulse arrays. For efficiency purposes when full pulses or an accurate energy spectrum are not needed, we have implemented a fast alternative to the full simulation. The ionization track is simulated to determine the timing of the event and the energy deposited in the gas. That energy is converted directly to an approximate Shaper measurement by smearing it with a Gaussian to roughly account for the missing physics and electronics response. The principal component of the NCD-phase analysis~\cite{ref:ncd_prl} that took advantage of the Shaper-only simulation was the determination of the neutron-detection efficiency.  On the other hand, the simulation of the alpha energy spectrum for the solar-neutrino analysis required the full pulse simulation.

\section{Impact and Implementation of the NCD Array Simulation}

\subsection{\label{sec:tuning}Tuning}

We tuned and validated the NCD array simulation by comparing simulated pulses to calibration data using a number of pulse-shape variables:
\begin{enumerate}
\item The physics model and detector response were tuned and validated by comparing $^{24}$Na neutron calibration data with $^{24}$Na neutron simulation~\cite{ref:24na}.
\item The string-to-string variation of the alpha contamination was tuned on data above the neutron-analysis energy window (above 1.0~MeV) and up to 7.0~MeV, where a pure alpha sample resides.
\item The alpha simulation, including systematic uncertainties, was validated in the region of interest for the NCD analysis by comparing it with calibration data from the alpha strings that were filled with $^4$He.
\end{enumerate}
This procedure was designed to ensure that the simulation accurately reproduces the data, without training on a data set that contains the NC neutron signal.

\begin{figure}
 \centering
\includegraphics[height=10.cm]{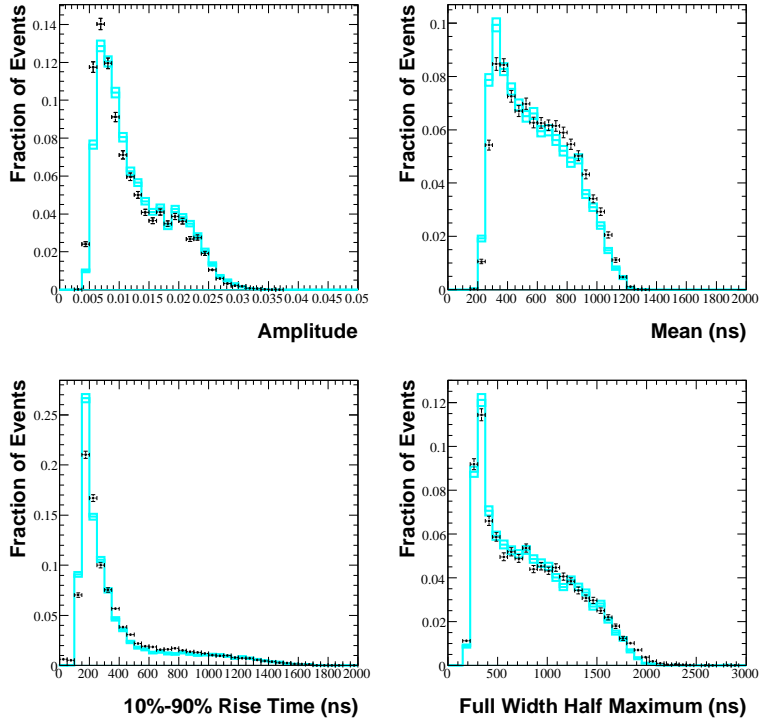}
\caption{\label{fig:neutron_psa} Comparison of pulse shape variables in $^{24}{\rm Na}$ neutron calibration data (black points) and the NCD array Monte Carlo (cyan curve) in the neutron energy window, 0.4 to 1.2~MeV, with statistical errors only.  From top left to bottom right: fraction of events as a function of pulse amplitude, time-axis mean of the pulse, 10\%-90\% rise time, and full width at half maximum.  All distributions are normalized to unit area.}
\end{figure}

Comparing the simulation to neutron calibration data tests nearly all aspects of the simulation physics model downstream of the primary energy loss by ions in the NCD counter gas. We compared simulated and real $^{24}$Na calibration data using the distributions of a number of pulse-shape-analysis variables. These included shape variables such as the pulse mean, width, skewness, kurtosis, amplitude, and integral; as well as timing variables, including various rise times (10\%-to-50\% and 10\%-to-90\% of the pulse amplitude), integral rise times, and full width at half maximum.  These comparisons were used to estimate parameter values and uncertainties for electron and ion motion in the NCD counter gas, as well as for the space charge model. After tuning, the agreement between data and simulation in these variables is generally good, as shown in \fref{fig:neutron_psa}. Using neutron calibration data to test the level of agreement with simulation is very valuable because it provides high statistics in the most relevant energy region for the analysis. Although the first analysis of the data from the NCD array~\cite{ref:ncd_prl} relied on energy and did not use pulse shape or timing variables, such comparisons build confidence that the simulation accurately models the development of signal pulses as a function of time, and that the physics is correctly modeled.

\begin{figure}
\centering
\includegraphics[height=10.cm]{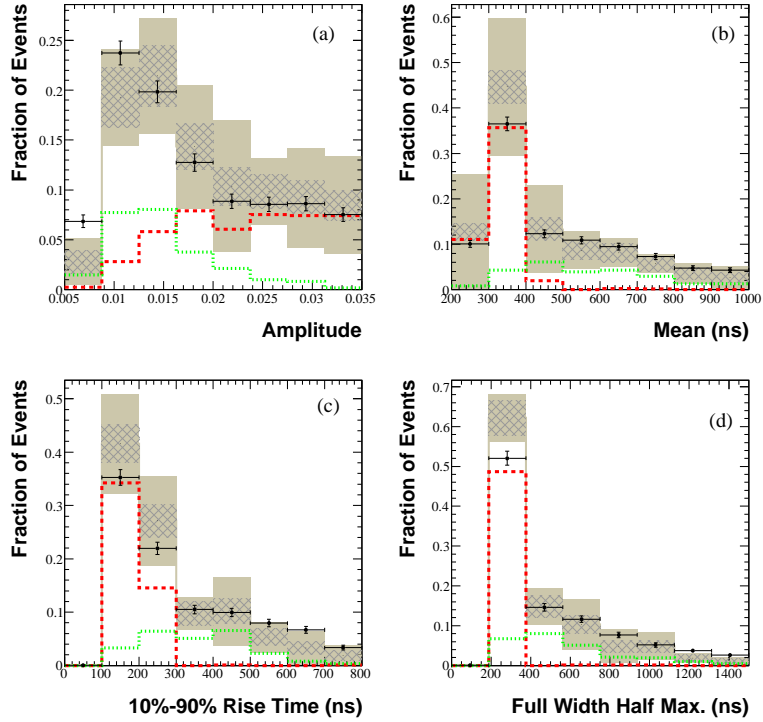}
\caption{\label{fig:alpha_psa} Comparison of pulse shape variables in $^{4}{\rm He}$ alpha calibration data (black points) and the NCD Monte Carlo alpha cocktail (grey curve), polonium alphas (red dashed line), and bulk alphas (green dotted line) in the NCD analysis energy window, 0.4 to 1.2 MeV.  The data points are shown with statistical errors, alpha cocktail Monte Carlo is shown with statistical (hatched) and systematic (grey filled) errors. From top left to bottom right: fraction of events vs. pulse amplitude, time-axis mean of the pulse, 10\%-90\% rise time, and full width at half maximum. The data and cocktail distributions are normalized to unit area, while the polonium- and bulk-alpha distributions are normalized to their fractional contribution to the alpha cocktail.}
\end{figure}

To compare data and Monte Carlo in regions of parameter space that contain significant numbers of alphas, it was necessary to construct a ``cocktail'' of simulated alpha pulses with the appropriate mixture of NCD cathode-surface polonium and bulk uranium and thorium alpha events.  This step was necessary because the pulse shape, timing, and energy distributions of polonium and bulk alpha events are rather different, as shown in \fref{fig:alpha_psa} and the right-hand plot of \fref{fig:energy_systematics} respectively.  Furthermore, the polonium-to-bulk ratio varies significantly from string to string, as does the total number of alpha events.  We estimated the fraction of polonium and bulk alpha events on each string by fitting each string's energy distribution in the region of 1.2 to 7.0~MeV with simulated polonium and bulk alpha-energy probability distribution functions (PDFs).  We chose to fit in energy for two reasons: (1) this variable provides excellent discrimination between surface polonium and bulk radioactivity, as can be seen in the right-hand plot of \fref{fig:energy_systematics}; (2) we found that energy is quite a robust variable against changes in Monte Carlo physics models. The latter point is not surprising since the energy depends on the total charge deposited in the counter. Therefore to first order it is independent of the details of the charge deposition process, unlike pulse-shape and timing variables.

Before the energy-spectrum fit, we calibrated the Monte Carlo energy for each string such that the peak of the polonium energy distribution in Monte Carlo matches that in data.  The size of the calibration constant (essentially an extra gain factor, applied as a multiplicative scale to the event energy) is typically 1-3\%, and is different for each string.  After the fit, we calculated an event weight, $w(string,\alpha_{type})$, which is a function of string number and alpha type (polonium, uranium, or thorium) describing the best-fit fraction of alphas on each string due to each source.\footnote{The bulk alpha events were assumed to be composed of equal parts uranium and thorium because their energy spectra are very similar. We were unable to measure the ratio of uranium and thorium on each string with sufficient accuracy because the normalizations of the independent uranium and thorium spectra were highly correlated in every energy fit.  A single combined bulk spectrum was used instead.}  In general, polonium comprised $\sim$60\% of the alpha population, however there were $\pm20\%$ (absolute) variations between strings.  The weights for the best-fit alpha fractions, and the data/MC energy scale correction, were applied on an event-by-event basis in the analysis.  

We validated the resulting cocktail alpha simulation by comparing with alpha data from the four $^4{\rm He}$-filled strings, from 0.4 to 1.2~MeV, and from the thirty-six $^3$He-filled strings, from 1.2 to 7~MeV. This comparison is shown in \fref{fig:alpha_psa} for several pulse-shape-analysis variables of interest. The agreement between the Monte Carlo and alpha calibration data is generally good, and builds confidence that the simulation physics model describes the alpha background well in the region of interest for the phase-three analysis.

\subsection{Identification of Alpha Backgrounds}\label{alphabackgrounds}
The fully-developed NCD array Monte Carlo improved our understanding of the alpha background events in the data.  The most common type of alpha event was due to a decaying nucleus in or on the nickel wall of a counter.  There should also be some radioactive contamination from the anode wire, although the characteristics of these pulses had not been well understood before the NCD Monte Carlo was developed.  There should also be alpha decays occurring in the endcap regions of the counters.  While an alpha travels in the gas in the endcap region, the electrons created as the alpha ionizes the gas do not reach the anode wire because of the silica feedthrough (see ~\cite{ref:ncd_nim} for details of the construction of the NCD counters).  Once the alpha reaches the region where the ionization electrons can drift to the anode wire, the drifting electrons may be affected by the locally-distorted electric field.  The shapes of the endcap alpha pulses were not well understood previously due to both of these effects.  The NCD array MC was used to characterize both types of minority alpha pulses.

The simulation of anode-wire-alpha pulses is a straightforward extension of standard ``wall-alpha'' pulses.  The origin of the alpha particle was set on or in the copper of the anode wire; the initial direction may be away from or into the wire (since alpha particles with enough energy can pass through the wire and still produce a pulse).  We simulated both bulk $^{238}$U and $^{232}$Th, and surface $^{210}$Po contaminations for the anode wires.

The simulation of endcap alpha pulses had the additional complication of tracking ionization electrons in the region where ionization electrons may not reach the anode wire, or where the electric field distortions may affect the pulse shapes.  Based on a rough calculation of the fields in the endcap regions we used a conical surface to describe the boundary where electrons start to drift to the anode.  Furthermore, we found that the field distortions were relatively small and could be ignored for the purpose of generally characterizing alpha pulses originating in the endcap regions.

\begin{figure}[htbp]
\begin{center}
\includegraphics[width=0.5\textwidth]{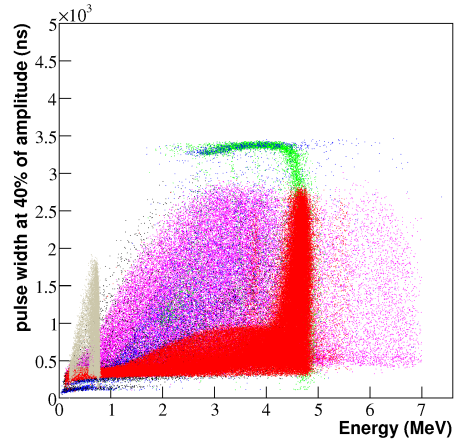}
\caption[Simulated Pulse-width versus Energy]{Pulse-width vs. energy distributions for simulated neutrons (grey) and alphas.  The alpha populations include surface wall $^{210}$Po alphas (red), bulk wall $^{238}$U alphas (magenta), surface wire $^{210}$Po alphas (green), bulk wire $^{238}$U alphas (blue), and bulk endcap $^{238}$U alphas (black, distributed with the blue points).}
\label{fig:widthvsemc}
\end{center}
\end{figure}

\begin{figure}[htbp]
\begin{center}
\includegraphics[width=0.5\textwidth]{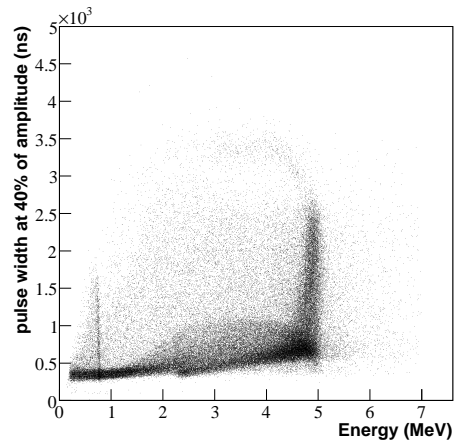}
\caption[Data Pulse-width versus Energy]{Pulse-width vs. energy distributions for the NCD array data.  Prior to studies with the NCD array MC the populations of extremely wide and narrow pulses were not well understood.  By comparing with the simulated data in \protect\fref{fig:widthvsemc} they can be identified as wire-alpha pulses and possibly endcap alpha pulses.}
\label{fig:widthvsedata}
\end{center}
\end{figure}

\Fref{fig:widthvsemc} shows the pulse-width vs. energy distribution for a variety of simulated alpha-pulse types and neutrons.  Besides the wall-alpha pulses there are also wire-alpha pulses and endcap alpha pulses.  An isolated population of wire-alpha events extends from the top of the $^{210}$Po-alpha peak, and some of the low-energy pulses are narrower than the wall alphas and neutrons.  Many of the endcap alpha pulses overlap with the wall alphas, but some of the low-energy endcap alphas are also narrower than the wall alphas and neutrons.  The pulse-width vs. energy distributions for simulated events in \fref{fig:widthvsemc} can be compared to the distribution for the NCD array data in \fref{fig:widthvsedata}.  There are clearly populations of events that match up with the wide wire-alpha pulses and with the narrow endcap alpha pulses.  Prior to the full development of the NCD array MC these populations in the data were not understood.  By simulating a variety of alpha pulse types we were able to identify these unexplained pulses in the data.

\subsection{Use in the Solar-Neutrino Signal Extraction}\label{useinanalysis}
In the SNO Phase-3 analysis~\cite{ref:ncd_prl} the number of events as a function of energy in the NCD array data were fit with neutron and alpha PDFs simultaneously with the PMT data.  The best-fit number of neutrons is proportional to the detected $^8$B solar-neutrino flux plus the deuteron photodisintigration background.  The neutron PDF comes from $^{24}$Na calibration data, while the alpha PDF is calculated with the NCD array simulation described in this paper.  Due to the lack of an adequate \emph{in-situ} alpha calibration, we needed the simulation to produce a data set with sufficient statistical accuracy and the correct bulk-to-surface event ratio.  We produced a data set with approximately 10-times the number of alpha events expected in the data from the SNO's third phase.  The final step was to determine the energy-dependent systematic uncertainties for the alpha energy spectrum.

\begin{table}[htb]
\caption{Parameters used in the NCD array simulation and their associated uncertainties, listed in descending order of importance.  The alpha cocktail fractions, the mean $^{238}$U \& $^{232}$Th depths, and the instrumental background cuts vary from string to string.}
\begin{indented}
\lineup
\item[]\begin{tabular}{@{}lll}
\br
Parameter                                        & Value         & Range             \\
\mr
Mean Po Depth ($\mu$m)                           & $\0\0{}0.1$   & $\pm \0\0{}0.1$   \\
Mean U \& Th Depth                               & Varies        & Varies            \\
Instrumental Background Cuts                     & Varies        & Varies            \\
Avalanche Width Gradient                         & $154$     & $\pm \0{}31$      \\
Avalanche Width Offset                           & $782$     & $\pm 120$         \\
Electron Drift Curve (Scaling, \%)               & $\0{}10$      & $\pm \0\0{}4$     \\
Ion Mobility (10$^{-8}$ cm$^2$ ns$^{-1}$ V$^{-1}$) & $\0\0{}1.082$ & $\pm \0\0{}0.027$ \\
Alpha Cocktail                                   & Varies        & Varies            \\
\br
\end{tabular}
\label{tab:paramstostudy}
\end{indented}
\end{table}

The systematic uncertainties we considered are listed in \tref{tab:paramstostudy}; these reflect the range of these parameters in the physics model of the NCD array.  Systematics are assessed by generating large sets of variational Monte Carlo data, each with one simulation input parameter varied by one $\sigma$ with respect to its default value, and with similar statistics to the standard Monte Carlo data set.  In the standard Monte Carlo data all parameters are fixed to their default values.  No other parameters, including energy scale and polonium/bulk fractions, are different between the default and variational Monte Carlo sets.  The size of the systematic-parameter variations for the NCD array Monte Carlo were estimated from \emph{ex-situ}, off-line measurements of the NCD array signal processing electronics response, and \emph{in-situ} constraints from the NCD array data.  In the latter case, we used orthogonal data sets to assess the variations, either from calibration data, or from the data set above 1.2~MeV.

The 1-$\sigma$ uncertainties for the selected systematic parameters were determined in different ways.  As described in \sref{sec:tuning}, the uncertainty on the alpha cocktail fractions comes from the high-energy fits. The alpha-depth uncertainties are conservatively assumed to be the difference between the best-fit values for the mean depths and completely uniform/surface distributions.  Qualitatively, varying the alpha depth changes the slope of the energy distribution in the neutron window because the width and the amplitude of the pulses depend on the origin of the alpha particles.  We fit the mean depths using alpha data above the neutron analysis window.  The uncertainty on the electron drift curve scaling is estimated from the analysis of the widest wire alphas, since this variation tends to change the average pulse width. The uncertainties on the space-charge offset and gradient parameters, which are the coefficients of $\cal{W}_s$ described in \sref{section:gain}, are determined by looking at how the features of the alpha energy spectrum shift as each parameter is changed.  Qualitatively, tuning these parameters changes the charge deposited versus time on the anode, and therefore the energy scale.  The ion mobility uncertainty is determined from the analysis of extremely narrow neutron calibration pulses.  The ion mobility variation affects the total amount of charge versus time as well.  The alpha fraction variation is also determined from the string-by-string fits described in \sref{sec:tuning}.  Qualitatively, varying these changes the slope of the alpha background energy distribution in the neutron energy window since the bulk and $^{210}$Po alphas have somewhat different slopes.

We propagated the Monte Carlo systematic uncertainties to the alpha energy distribution by calculating first derivatives~\cite{ref:Whalley:2002tu}.  For each of the eight systematic uncertainties, the first derivatives for each bin of the energy distribution were calculated by taking the difference between histograms containing the standard Monte Carlo prediction for the shape of the energy distribution, and the variational Monte Carlo energy-distribution shapes. Then, the total systematic uncertainty in each energy bin was assessed by summing the eight contributions in quadrature.  We calculate the uncertainty on the shape only (rather than shape and normalization) because the third-phase solar-neutrino analysis used an unconstrained fit for the neutron-signal and alpha-background normalizations.  We note that this procedure treats all systematic uncertainties as uncorrelated by fitting the variations as deviations.  

\begin{figure}
\centering
\mbox{
\begin{minipage}{0.32\textwidth}
\includegraphics[width=\textwidth]{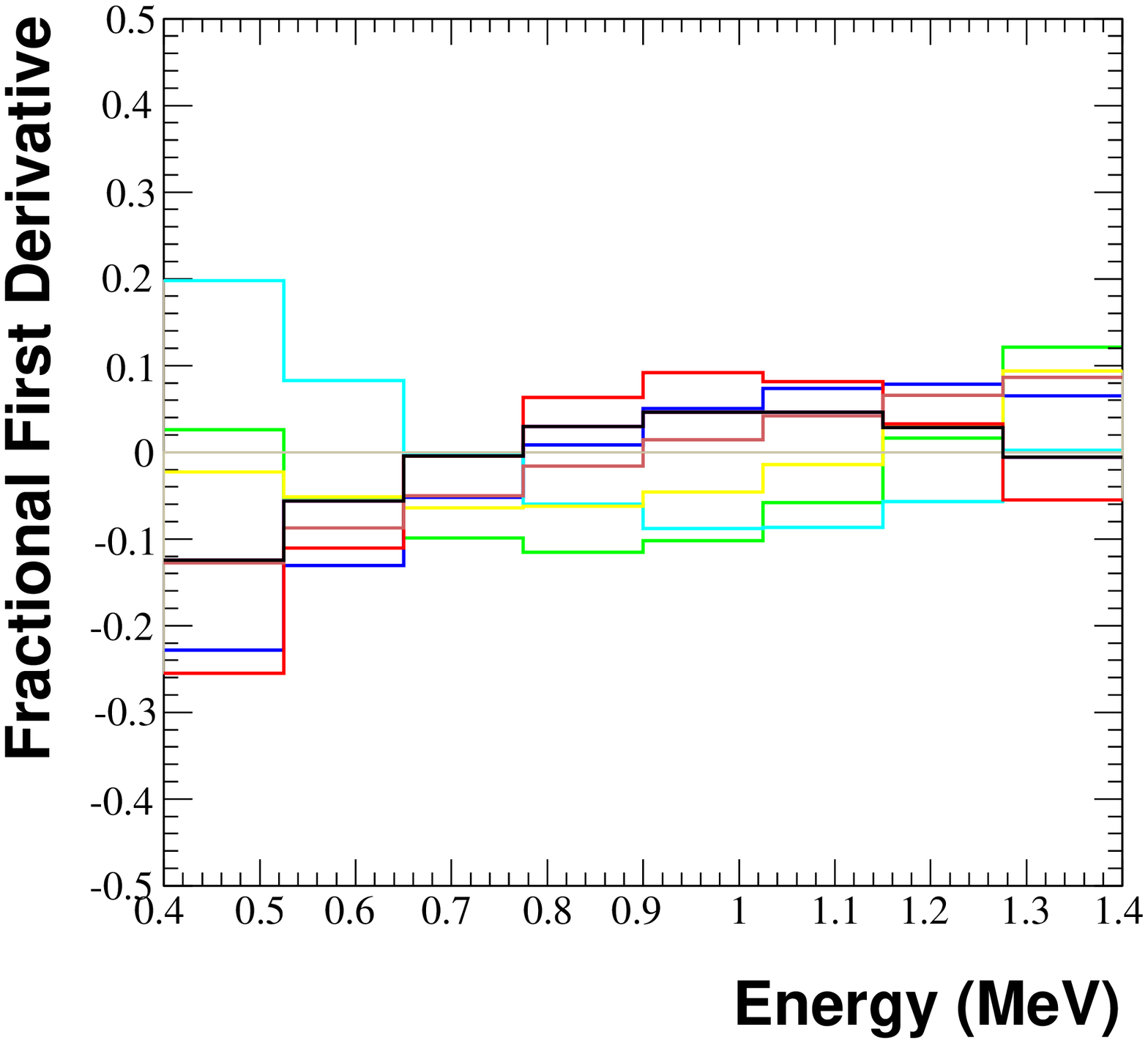}
\end{minipage}
\begin{minipage}{0.32\textwidth}
\includegraphics[width=\textwidth]{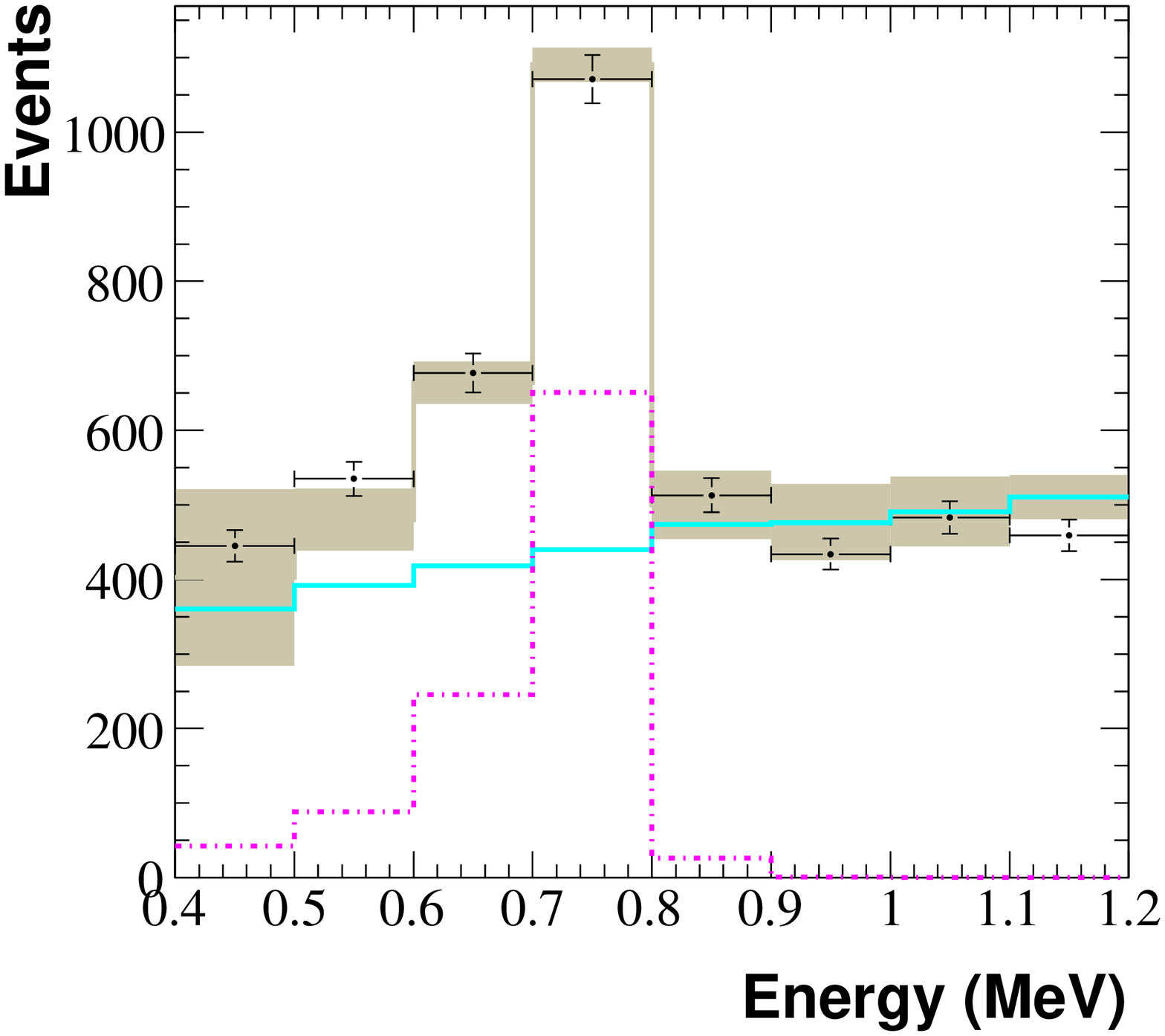}
\end{minipage}
\begin{minipage}{0.32\textwidth}
\includegraphics[width=\textwidth]{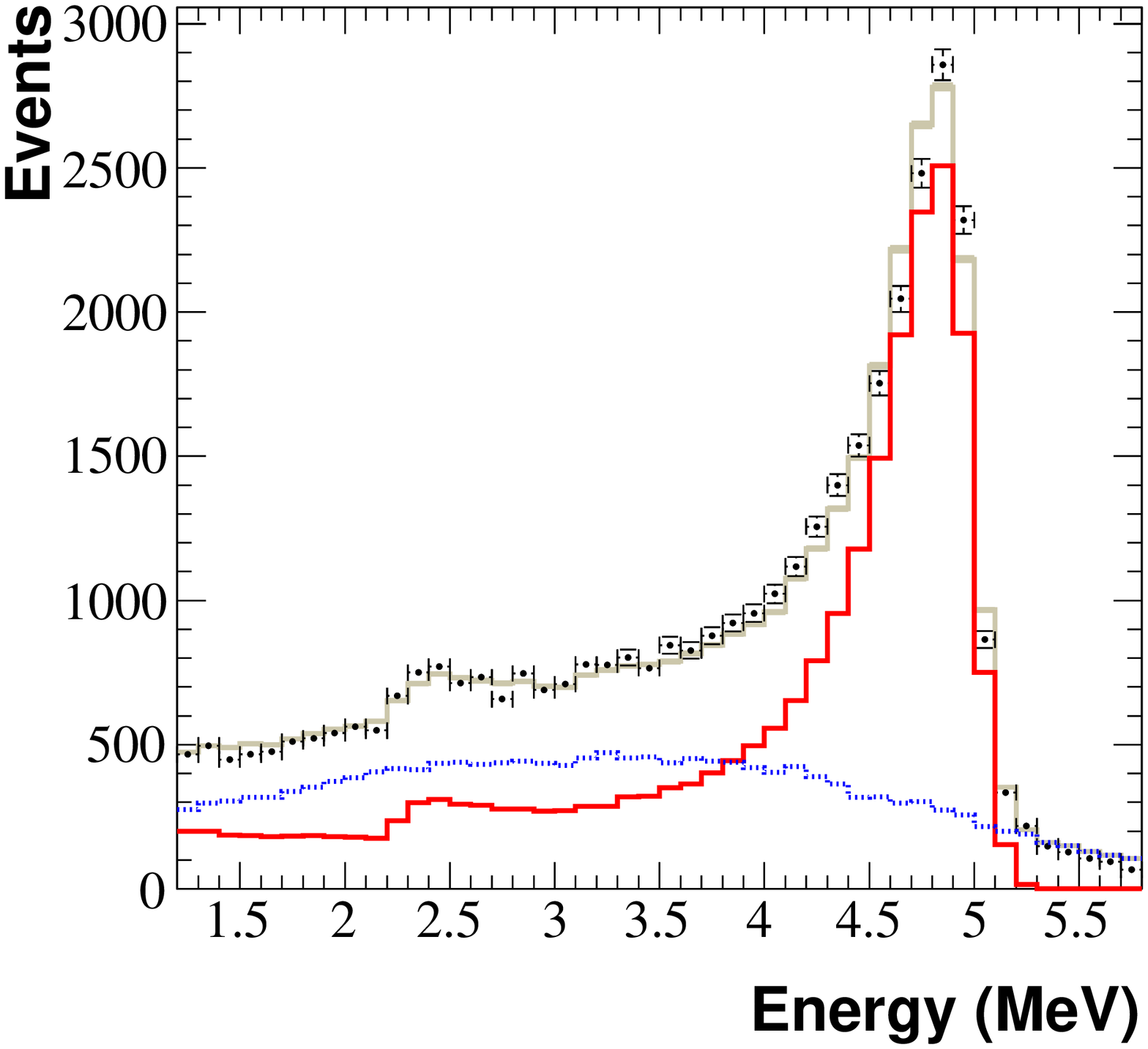}
\end{minipage}
}
\caption{Left: fractional first derivatives vs. energy (MeV) for significant systematic variations of the simulation. Center: number of events vs. energy (MeV) in the NCD analysis energy window, for data (black points, with statistical uncertainties), neutron PDF from calibration data (purple dashed curve), alpha cocktail PDF from simulation (cyan curve), and total predicted number of events (grey curve, with systematic uncertainties).  Right: number of events vs. energy (MeV) above the NCD array analysis energy window, for data (black points, with statistical uncertainties), alpha cocktail (wide grey curve, with systematic uncertainties), polonium (red curve), and bulk (blue dashed curve) simulation.}
\label{fig:energy_systematics}
\end{figure}

To incorporate the simulation systematic uncertainties, the solar-neutrino analysis of the third phase used the alpha background first derivatives described above. For ease of use, we generated analytical functions describing the fractional first derivatives as a function of energy, which are shown in the left-hand plot of \fref{fig:energy_systematics}. The best fit functions were determined by $\chi^2$ minimization to be second-order polynomials.  These functions multiply the standard Monte Carlo energy distribution to produce the one-$\sigma$ variational background PDFs.  In the third-phase analysis, each function was multiplied by a nuisance parameter for the normalization of that particular uncertainty.  The nuisance parameters were constrained with a penalty term in the likelihood function.

To estimate the impact of these systematic uncertainties we did two kinds of tests. First, we estimated the errors on the pulse shape and timing distributions to give guidance on which variables are most robust against NCD array Monte Carlo physics model uncertainties. We found that some variables were much more sensitive to certain systematic variations than others. For example, the space-charge model parameters primarily affect the rise-time distributions, while the alpha-depth variations have a large effect on most variables. Second, we fit the energy spectrum of an unknown fraction of the data (to maintain blindness) to determine the number of neutrons between 0.4 to 1.2~MeV, using each of the variational Monte Carlo sets as the alpha-background PDF. The variation in the number of neutrons provided the most relevant estimate of the uncertainty, and possible bias, due to each systematic source. In general, these fits showed that the alpha depth variation systematic uncertainty has the largest impact, followed by the noise model, the electron drift time curve, and the ion mobility variation. The total systematic uncertainty assessed in this way on the number of neutrons was approximately 4\%.

The final simulated alpha energy spectrum in the analysis energy window is shown in the center plot of \fref{fig:energy_systematics}, with systematic uncertainties. This alpha spectrum was used as the background PDF in the third-phase analysis, together with the neutron signal PDF from $^{24}$Na neutron calibration data, to extract the best-fit total number of neutron events.

\subsection{Pulse Fitting}\label{sec:pulsefitting}
Another approach to separating neutron and alpha events is to fit individual pulses with neutron and alpha template pulses.  In particular, we can use the NCD array Monte Carlo to generate libraries of such templates for both the signal neutrons and alpha backgrounds.\footnote{It would be inefficient to generate the pulses as a fit is performed because of the time needed to generate each pulse.  Furthermore, most minimizing routines perform best when the fit function (i.e. the simulated pulse shape) changes smoothly as a function of its parameters ($r$, $E$, $\theta$, $\phi$, and $z$).  This is not the case in the NCD array simulation when a change of any of those coordinates can cause a track to hit the wall or wire and therefore change the pulse shape discontinuously.}  A library of simulated pulses consists a set of a single class of pulses (either neutrons, wall alphas, or wire alphas), where the pulse shapes included cover as many of the available pulse shapes as possible.  The parameter space for the available pulse shapes consists of five variables: the radius, $r$, the two track angles, $\theta$ and $\phi$, the position along the NCD string, $z$, and the initial energy, $E$.  A data pulse is fit by finding the best match in the library.

The three types of events, neutrons, wall alphas, and wire alphas, are catalogued in separate libraries.  The three libraries are generated by varying the parameters on the four-dimensional grids in their respective parameter spaces.  For neutrons the relevant coordinates of the parameter space are $r$, $\theta$, $\phi$, and $z$.  The initial energy is fixed at 574~keV and 191~keV for the proton and triton, respectively.  For alphas the relevant coordinates are $E$, $\theta$, $\phi$, and $z$, with the radius fixed at either the inner radius of the wall or the outer radius of the wire (the pulse shape for any alpha leaving the wall with a given $E$, $\theta$, $\phi$, and $z$ will be the same regardless of whether or not it started out deeper in the nickel or on the surface).  \Tref{tab:librarygrids} shows the grids used in the pulse-shape parameter spaces, and \tref{tab:librarysizes} shows the numbers of pulses in the three libraries.  Further details about the creation of the pulse libraries can be found in~\cite{ref:nsothesis}.

\begin{table}[htdp]
\caption[Pulse Library Grid Attributes]{The number of grid points along each dimension in the relevant simulation parameter space for the three libraries produced.  The radius points are spaced uniformly in $r$.}
\begin{indented}
\lineup
\item[]\begin{tabular}{@{}lccccc}
\br
 & $r$ (cm) & $E$$^{\mathrm{a}}$ (MeV) & $\theta$ ($^{\circ}$) & $\phi$$^{\mathrm{b}}$ ($^{\circ}$) & $z$ (cm) \\
\mr
Range & $[0.1,\;2.5]$ & $[0.2,\;1.2]$ & $[0,\;90]$ & $^{\mathrm{c}}$ & $[-535,\;535]$ \\
\mr
Neutron    & 5   & N/A   & 10 & 10 & 10 \\
Wall Alpha & N/A & 10+2$^{\mathrm{d}}$ & 10 & 5  & 10 \\
Wire Alpha & N/A & 10    & 10 & 5  & 10 \\
\br
\end{tabular}
\item[]$^{\mathrm{a}}$ This energy is the initial energy of the alpha particle.
\item[]$^{\mathrm{b}}$ The number of $\phi$ points depends on $\theta$ to uniformly cover the spherical space.  The maximum number of $\phi$ points is given in this table.
\item[]$^{\mathrm{c}}$ The specific $\phi$ range used depended on the geometry relevant for each library.
\item[]$^{\mathrm{d}}$ Two extra grid points are added in energy to account for high-energy ions that reenter the NCD wall, depositing only some of their energy in the gas.
\end{indented}
\label{tab:librarygrids}
\end{table}

\begin{table}[htdp]
\caption[Pulse Library Sizes]{The number of pulses generated and used for the neutron, wall-alpha, and wire-alpha libraries.  The number of pulses generated is determined by the parameter-space grids given in \protect\tref{tab:librarygrids}.  Some pulses were removed because their deposited energies were either too high or too low.  Pulses in the energy range $0.2 < E < 1.2$~MeV were kept.}
\begin{indented}
\lineup
\item[]\begin{tabular}{@{}lccc}
\br
 & Neutron & Wall Alpha & Wire Alpha \\
\mr
Generated & 3350 & 4200 & 3500 \\
Used      & 3329 & 2974 & 2599 \\
\br
\end{tabular}
\end{indented}
\label{tab:librarysizes}
\end{table}

When comparing a data pulse to a library of simulated pulses, the quality of each fit is determined by a Pearson's $\chi^2$ test.  The library pulse that has the smallest $\chi^2$ per degree of freedom (dof) when compared to the data pulse is the best fit to the data.

The fit region for each pulse is determined by finding where the pulse amplitude drops to a given fraction of the peak amplitude on the rising and falling edges.  The tail of each neutron-capture and alpha pulse quickly becomes dominated by the characteristic ion-drift time constants, while the rising edge contains information about the particle that produced the pulse.  As a result, the left (rising) and right (falling) edges of the pulse are treated differently.  The left edge of the fit region is the point at which the pulse crosses 10\% of its peak value.  10\% of the peak amplitude is typically above the baseline noise, even for low pulses, and almost the entire rise of the pulse is included in the fit region.  The right edge of the fit region is the point at which the pulse crosses 30\% of its peak value.  A larger percentage is chosen for the falling edge than the rising edge to avoid most of the ion-drift tail, which is almost the same for every pulse and would therefore make the $\chi^2$ parameter less effective in separating pulse shapes.

The variance used in the $\chi^2$ calculation is also determined for each pulse.  There are four sources of variance that can contribute to the difference between a MC library pulse and a data pulse:
\begin{itemize}
\item Electronic noise: Various parts of the electronics contribute to the background noise on a pulse.  The exact frequency spectrum of that noise depends on the bandwidths of downstream electronics components, but it can be approximated by taking the RMS of the tail of a pulse about the mean value.  In particular, bins 11,000 to 14,999 (from approximately 10.2 to 14.2~$\mu$s after the start of the pulse) are used.  This contribution to the variance is the largest of the four.
\item Digitization: When the pulse is digitized some uncertainty is added to every bin because each digitized value could represent a range of actual values.  The non-voltage-dependent portion of the digitization variance is already accounted for by taking the RMS of the tail of the pulse.  The voltage-dependent portion is calculated separately.
\item Library sparseness: Since the MC libraries are created on grids in the pulse parameter spaces before any fitting is performed, they represent a selection of the available pulse shapes.  Differences between the best-fit MC library pulse and the actual track parameters of the pulse being fit will add to the variance.
\item MC imperfections: Since the MC is not perfect, even if an MC pulse is created with the exact same track parameters as a real pulse there will still be differences between the two pulses because the MC is not an exact model.
\end{itemize}

The digitization contribution to the variance depends on the size of a single step in the digitizer, $\Delta$:
\begin{equation}\label{eq:sheppard}
\sigma_{\mathrm{D}}^2 = \frac{\Delta^2}{12}.
\end{equation}
This factor, known as ``Sheppard's Correction,''~\cite{ref:sheppardscorrection} is the correction that would need to be applied to determine the true width of a Gaussian peak in a spectrum that has been binned.  In other words, it is the variance of a uniform distribution, since the distribution is uniform within a bin.  The logarithmic amplification of the NCD arraypulses makes $\Delta$ voltage dependent.  It is also dependent on the characteristics of the logarithmic amplifiers that go into the analytic description of the amplification, \eref{eq:logamp}.  A change of input voltage by $\Delta$ results in a change of $V_{\mathrm{log}}$ by 1 digitization unit:
\begin{equation}
V_{\mathrm{log}}(t) + 1 = a \log_{10}\left(1 + \frac{V_{\mathrm{lin}}(t-\Delta{}t) + \Delta}{b} \right) + c_{\mathrm{chan}} + V_{\mathrm{PreTrig}}.
\end{equation}
Solving for $\Delta$, and substituting into~\eref{eq:sheppard}, the contribution to the variance due to digitization is:
\begin{equation}\label{eq:deltaderiv3}
\sigma_{D}^2 = \frac{1}{12}\left(\frac{V_{\mathrm{lin}}(t-\Delta{}t) + b}{a\log_{10}e}\right)^2.
\end{equation}
\Eref{eq:deltaderiv3} can be expanded and separated into two voltage-dependent terms, and a voltage-independent term.  The latter is a part of the variance that is calculated by taking the RMS of the noise in the tail of each pulse.  The voltage-dependent component of the digitization variance turns out to be a relatively small contribution to the total variance.  It is an order of magnitude smaller than the voltage-independent variance (including the electronics and part of the digitization), for pulses with large amplitudes (and therefore the largest digitization variance).  

Unfortunately there is no well-defined method for quantifying the contributions to the variance from the sparseness of the libraries or the imperfections in the MC.  We studied the increase in the mean reduced $\chi^2$ as a result of these two contributions by fitting simulated and real $^{24}$Na neutron pulses with the neutron library.  The library was used to fit a simulated $^{24}$Na neutron data set; the excess $\chi^2/dof$ above $1$ was a result of the variance contribution from the library sparseness.  The library was then used to fit a real $^{24}$Na neutron data set, and the excess $\chi^2/dof$ above the value from fitting the simulated data was a result of the variance contribution from MC imperfections.  We found that their contributions were small and had no significant impact on the ability to separate neutron and alpha pulses.

An energy term of the form $(E-E_{\mathrm{fit}})^2/V[E]$ is also added to the $\chi^2$ statistic to take into account the difference in energy between the library pulse and the pulse being fit.  The variance for the energy term is set by the alpha-library energy spacing, $\Delta E_{\alpha}$: $V[E] \equiv (\Delta E_{\alpha}/2)^2$.  For the libraries used, $\Delta E_{\alpha} = 0.1 \mathrm{MeV}$.  This factor has a small effect on the total $\chi^2$ when the difference between the energies is small, particularly since the number of bins in each pulse's fit region is large.

\begin{figure}[htbp]
\begin{center}
\includegraphics[width=0.45\textwidth]{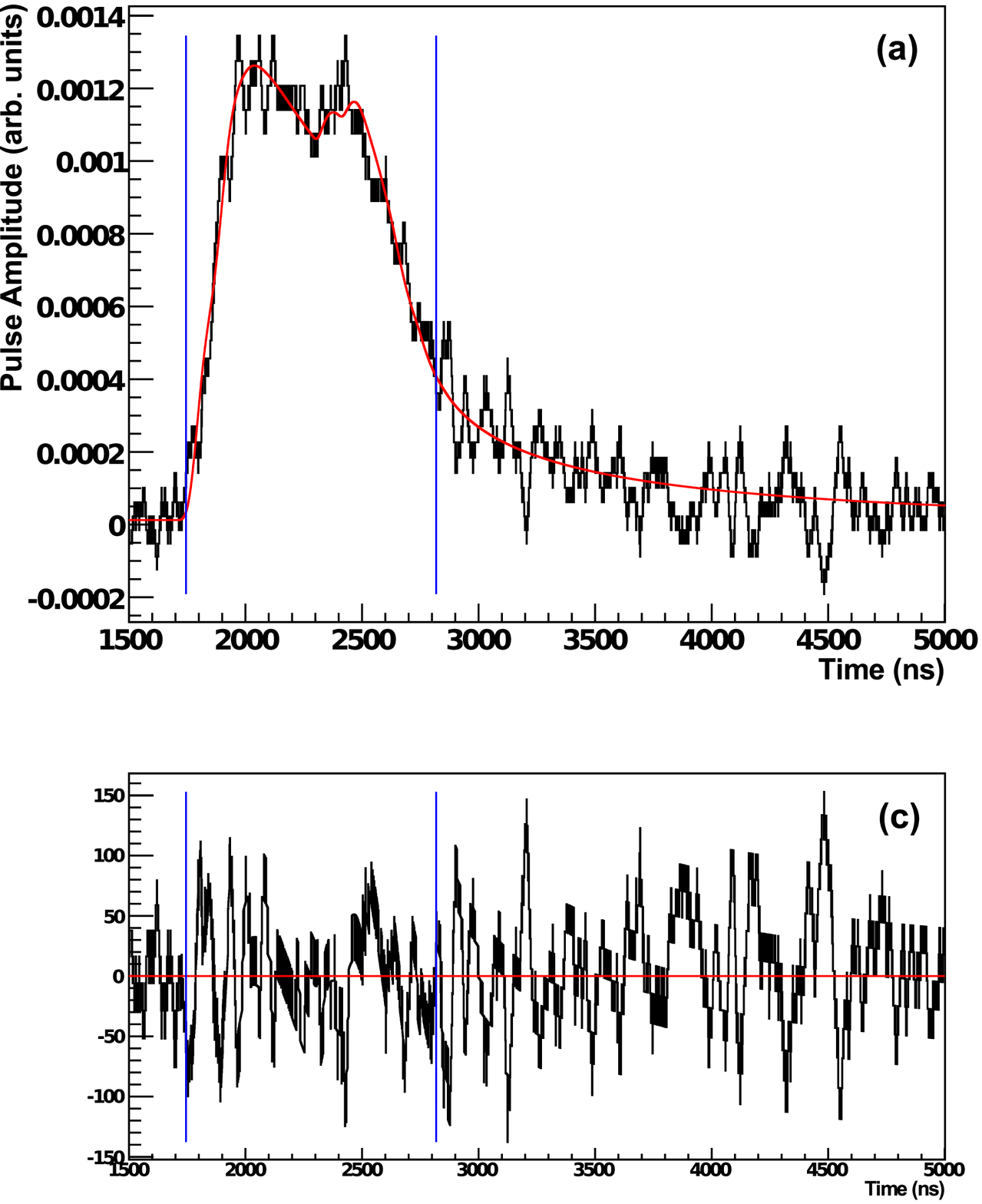}
\includegraphics[width=0.45\textwidth]{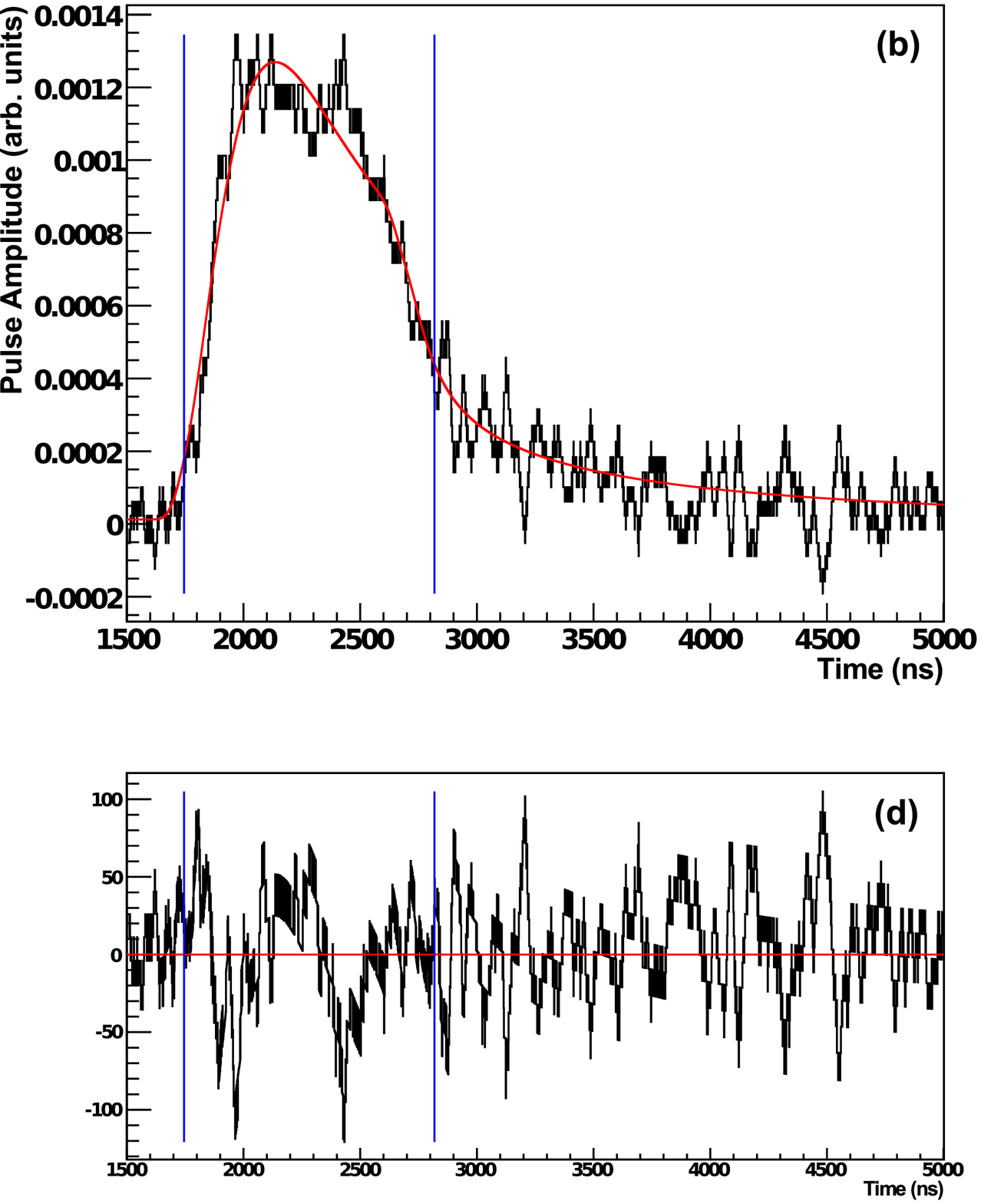}
\caption[Neutron fit example]{An example of a Monte Carlo neutron pulse fit with the neutron library (a), and the same pulse fit with the alpha library (b).  The vertical blue lines indicate the fit region, in which the $\chi^2$ was calculated.  The $\chi^2$ in (b) is larger than in (a), though this example serves to show how a neutron pulse (even one that is not extremely narrow) can still be fit fairly well with the alpha library.  Plots (c) and (d) are the respective fit residuals.}
\label{fig:nfitexample}
\end{center}
\end{figure}

\begin{figure}[htbp]
\begin{center}
\includegraphics[width=0.45\textwidth]{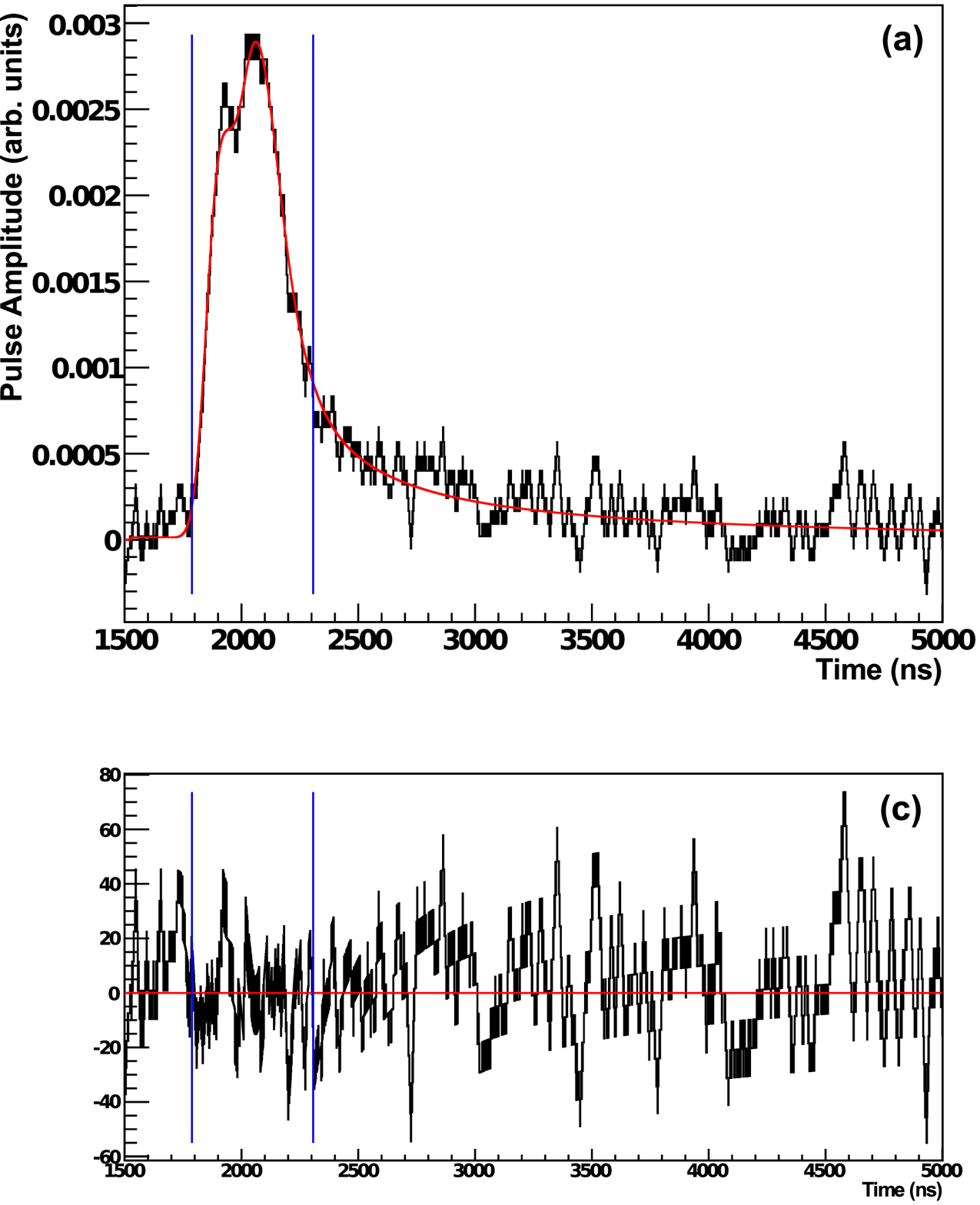}
\includegraphics[width=0.45\textwidth]{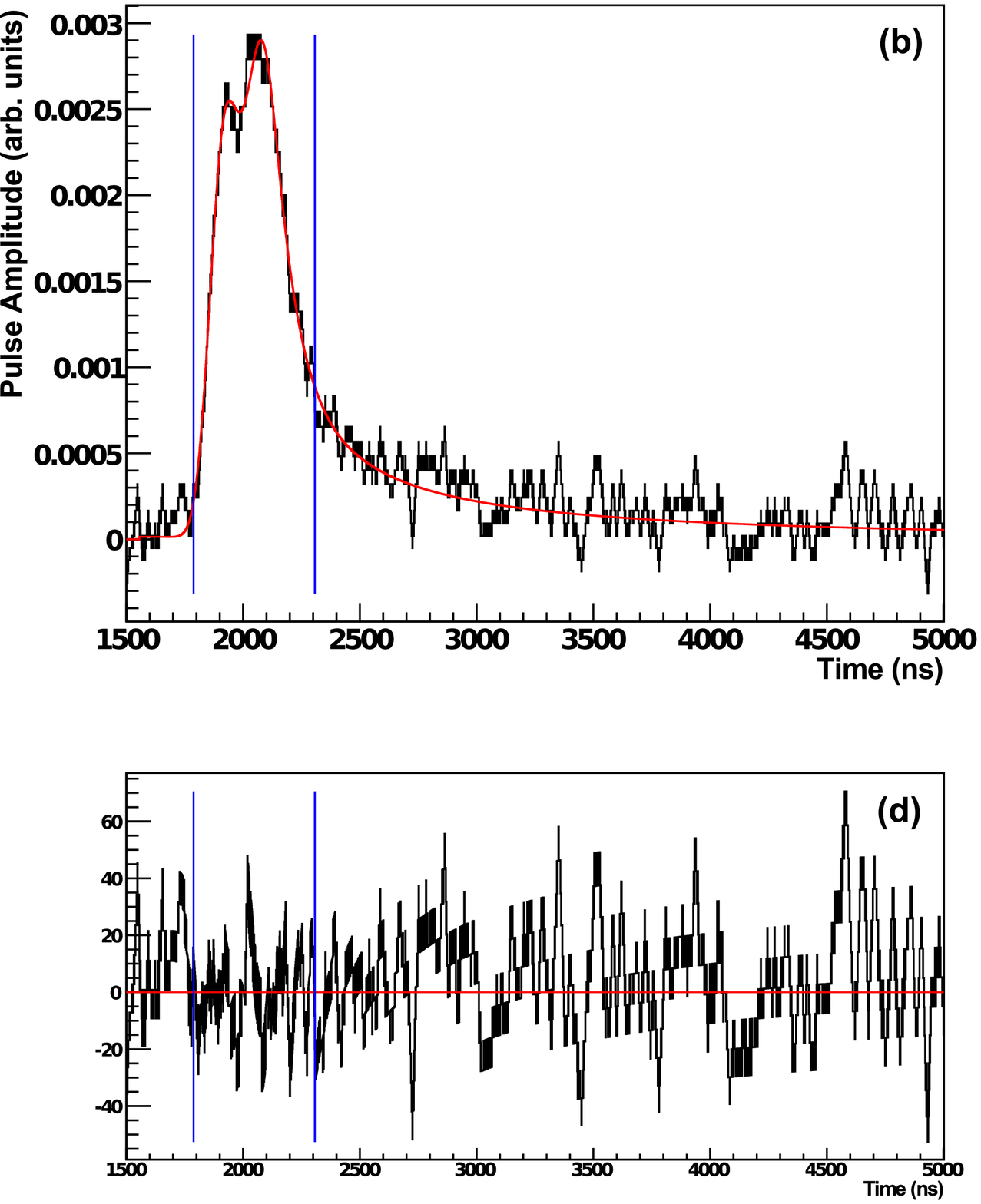}
\caption[Alpha fit example]{An example of a Monte Carlo alpha pulse fit with the neutron library (a), and the same pulse fit with the alpha library (b).  The vertical blue lines indicate the fit region, in which the $\chi^2$ was calculated.  Alphas tend to be fit well with both the neutron and alpha libraries, as is the case with this example.  Plots (c) and (d) are the respective fit residuals.}
\label{fig:afitexample}
\end{center}
\end{figure}

We first fit simulated data with the pulse libraries.  \Fref{fig:nfitexample} shows an example Monte Carlo neutron pulse fit with both the neutron library and the alpha library (the latter is a combination of the wall- and wire-alpha libraries), accompanied by the respective fit residuals.  \Fref{fig:afitexample} shows a fit of an alpha Monte Carlo pulse with the neutron and alpha libraries.  The neutron pulse selected is fit well with the neutron library ($\chi^2/dof = 0.58$, with approximately 1070 degrees of freedom) and fairly well with the alpha library ($\chi^2/dof = 0.97$).  Qualitatively, it is clear that the fit with the neutron library better represents the overall structure of the pulse.  Some neutrons have a more generic pulse shape and are fit reasonably well with both the neutron and alpha libraries.  The alpha pulse, as is the case with most alpha pulses, fits well with both libraries ($\chi^2/dof = 0.267$ for both fits).

\begin{figure}[htbp]
\begin{center}
\includegraphics[width=0.45\textwidth]{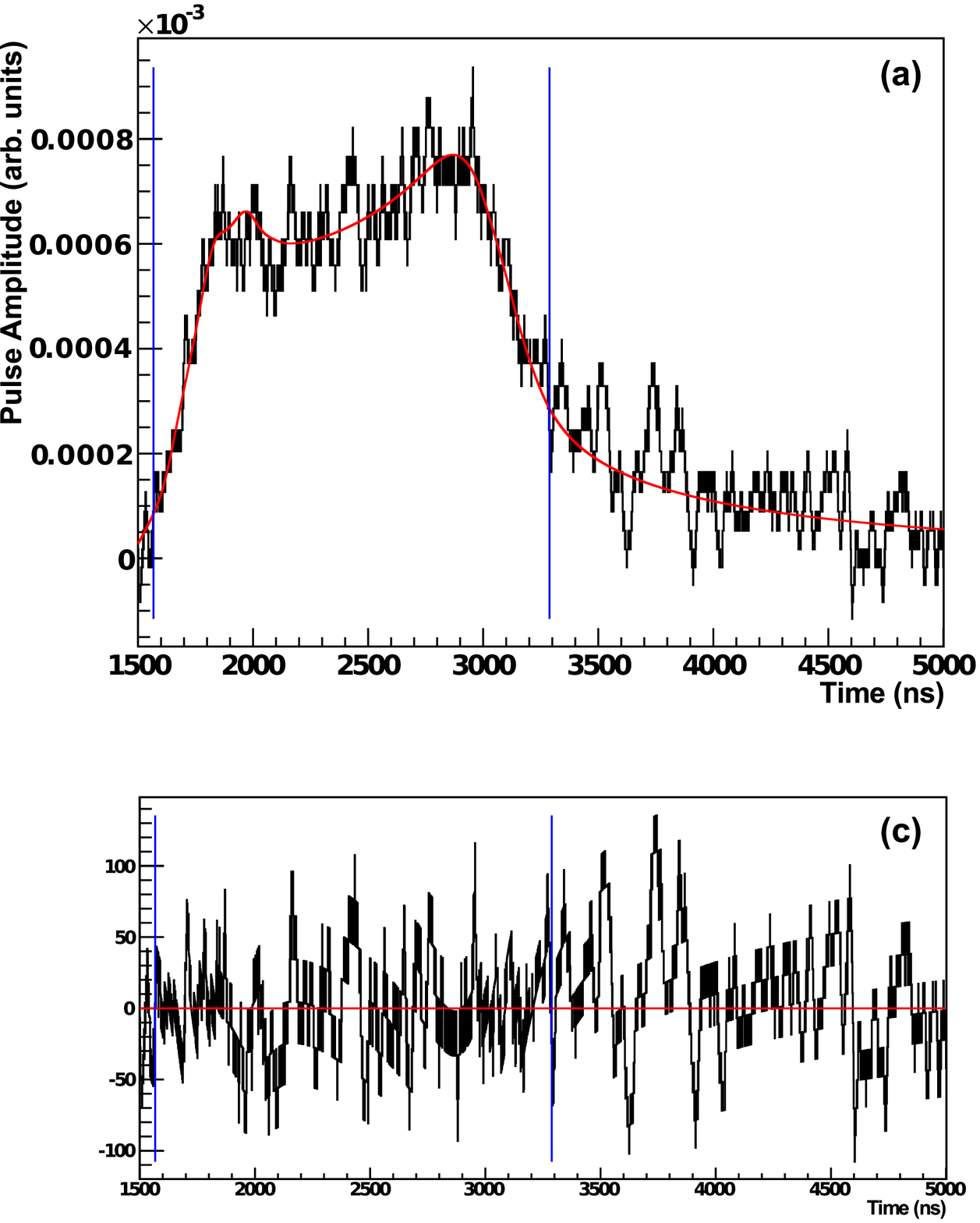}
\includegraphics[width=0.45\textwidth]{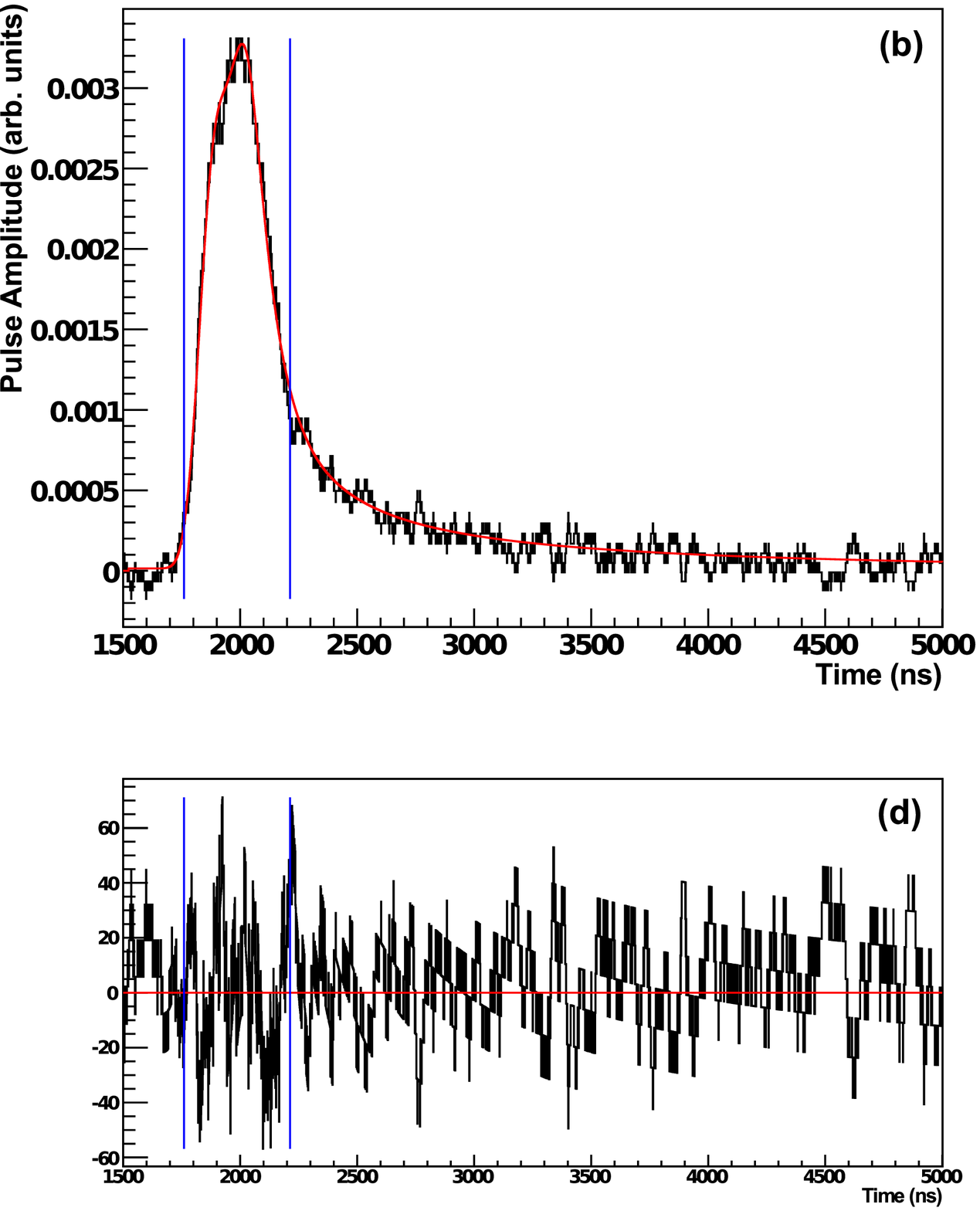}
\caption[Data fit examples]{(a) A pulse from the $^{24}$Na data set fit with the neutron library and (b) a pulse from the $^4$He strings fit with the alpha library.  The blue lines indicate the fit region, in which the $\chi^2$ was calculated.  Plots (c) and (d) are the respective fit residuals.}
\label{fig:datafitexamples}
\end{center}
\end{figure}

Fitting calibration neutrons and $^4$He alpha pulses with the simulated-pulse libraries is a strong test of the data-MC agreement.  \Fref{fig:datafitexamples} shows two examples: (a) a $^{24}$Na neutron pulse and (b) a $^4$He alpha pulse fit with the neutron and alpha libraries, respectively ((a) $\chi^2/dof = 0.319$; (b) $\chi^2/dof = 1.091$).  In general, these fits work well.

\begin{figure}[htbp]
\begin{center}
\includegraphics[width=0.8\textwidth]{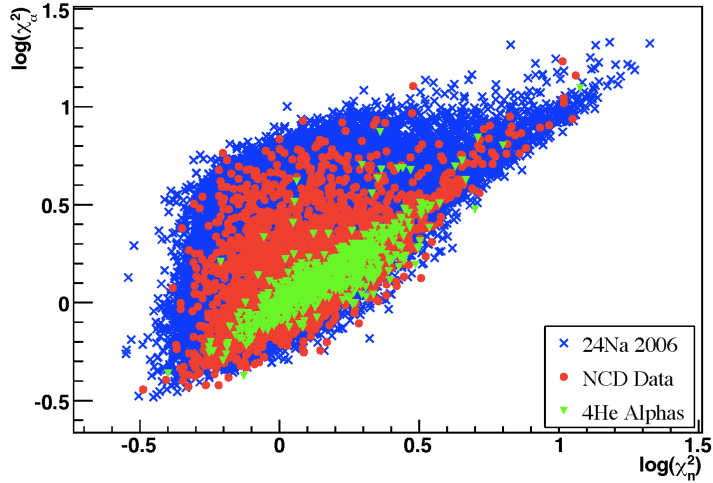}
\caption[$\chi^2_{\alpha}$ versus $\chi^2_n$]{Comparison of the fit results for fitting neutrons and alphas with the simulated neutron and alpha pulse libraries.  The axes are $\log\chi^2_{\alpha}$ versus $\log\chi^2_n$.  The neutrons are from $^{24}$Na calibrations, and the alphas are from the $^4$He strings.  The full NCD-phase data set is also plotted.  The $^{24}$Na calibration data are comprised of approximately 36-times the expected number of neutrons in the phase-three data, and the $^4$He alpha data are comprised of approximately half as many alphas expected in the NCD-phase data.  There is a subset of the neutron pulses which can be selected with little contamination from alpha pulses.}
\label{fig:dlrcs2d}
\end{center}
\end{figure}

The fit results can be used to separate neutron and alpha events because low-energy alpha pulses ($< 1$~MeV) resemble neutron pulses, but a significant fraction of the neutron pulses look distinctly different than any alpha pulse.  The comparison of the fit results, $\chi^2_{\alpha}$ versus $\chi^2_n$, where $\chi^2_{\alpha}$ is the result of the fit with the alpha library, and $\chi^2_n$ is the result of the fit with the neutron library, is an effective discriminator of different pulse classes.  \Fref{fig:dlrcs2d} shows the distributions of $\log\chi^2_{\alpha}$ versus $\log\chi^2_n$ for calibration neutrons, $^4$He-string alphas, and the NCD-phase data set.  The alphas are grouped along the $\chi^2_{\alpha} = \chi^2_n$ line, as are some of the neutrons.  There is a large group of neutron pulses that fit better with the neutron library than with the alpha library.

\begin{figure}[htbp]
\begin{center}
\includegraphics[width=0.8\textwidth]{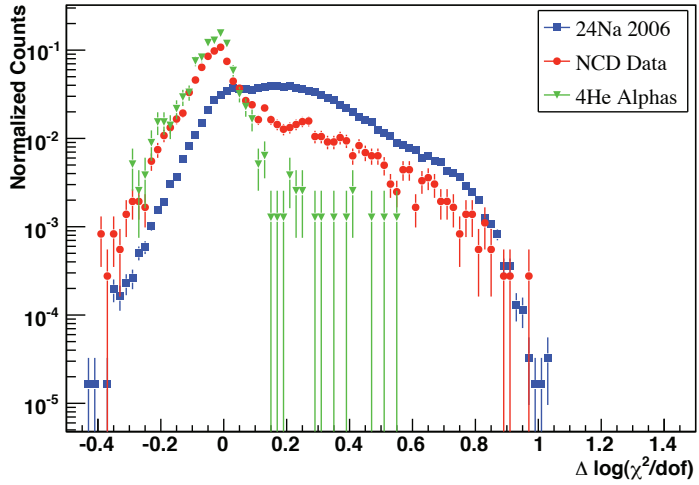}
\caption[$\log(\chi^2_{\alpha}) - \log(\chi^2_n)$]{The fit results from the previous figure projected along the $\Delta\log(\chi^2) \equiv \log(\chi^2_{\alpha}) - \log(\chi^2_n)$ axis. The wide distribution of neutron pulses on the right can be selected with little alpha background.  The histograms are normalized to unit area.}
\label{fig:dlrcs1d}
\end{center}
\end{figure}

The one-dimensional projection $\Delta\log(\chi^2) \equiv \log(\chi^2_{\alpha}) - \log(\chi^2_n)$ is more useful for seeing the ability to select a sample of neutrons with almost no alpha background.  This projection is shown in \fref{fig:dlrcs1d}.  The peak on the left is due to alpha and alpha-like neutron pulses.  The wider distribution of neutrons on the right is the sample of neutron pulses that are easily distinguished from alpha pulses.  These neutron pulses can be selected and analyzed with little background alpha contamination.  This type of pulse-shape analysis has the potential to increase the signal-to-noise ratio in the extraction of the number of neutrons detected in future analyses of SNO phase-three data.  The details of such an analysis are beyond the scope of this paper, but can be found in~\cite{ref:nsothesis}. The simulation-based pulse-shape analysis is being considered for use in the final SNO three-phase analysis~\cite{ref:psaposter}.

\clearpage

\section{Conclusions}
We have developed a unique and detailed simulation of the current pulses from the NCD proportional counters.  The simulation model includes energy loss by particles traveling through the NCD counter gas, walls, endcaps, and wire, as well as energy straggling.  In the gas, the simulation includes ion drift and electron diffusion.  Near the anode wire, the simulation models avalanche multiplication and the effects of space charge. The signal processing chain model includes the full NCD array data-acquisition electronics response, and its associated frequency-dependent noise.

The simulation accurately models the current pulse shapes collectively and on a pulse-by-pulse basis, for both signal neutrons and background alphas. First principles calculations are used for the majority of the simulation physics.  Some model parameters are measured with calibration data, and others are parameterized to match neutron calibration and high-energy alpha data. The dominant systematic uncertainties on the simulation physics come from the alpha implantation depth in the NCD counter walls, the electronics noise model, the electron drift time curve, and the ion mobility variation. 

The NCD array Monte Carlo simulation has been used to great effect in identifying classes of alpha backgrounds, in modeling the energy spectrum for the total alpha background in the third-phase $^8$B solar-neutrino measurement~\cite{ref:ncd_prl}, and in pulse fitting to help identify neutron and alpha pulses.

\end{document}